\documentclass[aps,prb,twocolumn,superscriptaddress,longbibliography]{revtex4-2}
\usepackage{newunicodechar}
\newunicodechar{−}{-} 
\usepackage{amsmath}
\usepackage{amssymb}
\usepackage{hyperref}
\usepackage{graphicx}         
\usepackage{dcolumn}          
\usepackage{bm}               
\usepackage{multirow}
\usepackage{epstopdf}
\usepackage{color}
\usepackage{gensymb}
\usepackage{textcomp}
\usepackage{comment}
\usepackage{chemformula}      

\begin{document}

\title{Competing Pair Density Wave and Uniform $d$-wave Superconductivity \\ 
in Phase Separated 214 Cuprates at the 1/8 Anomaly}


\author{Q.~Chen}
\affiliation{%
 Department of Physics and Astronomy, McMaster University, Hamilton, Ontario, L8S 4M1, Canada
}%
\affiliation{%
 Brockhouse Institute for Materials Research, Hamilton, Ontario, L8S 4M1, Canada
}%

\author{A. Moskal}
\affiliation{%
 Department of Physics and Astronomy, McMaster University, Hamilton, Ontario, L8S 4M1, Canada
}%

\author{Y. Wang}
\affiliation{%
 Department of Physics and Astronomy, McMaster University, Hamilton, Ontario, L8S 4M1, Canada
}%
\affiliation{%
 School of the Gifted Young, University of Science and Technology of China, Hefei, China
}%

\author{B. D. E. McNiven}
\thanks{Present Address: Department of Electrical and Computer Engineering, Memorial University of Newfoundland, St. John’s, Newfoundland and Labrador, Canada}
\affiliation{%
 Department of Physics and Physical Oceanography, Memorial University of Newfoundland, St. John’s, Newfoundland and Labrador, Canada A1B 3X7
}%

\author{A. A. Aczel}
\author{W. Tian}
\affiliation{%
    Neutron Scattering Division, Oak Ridge National Laboratory, Oak Ridge, TN 37831, USA
}%

\author{B. D. Gaulin}%
\affiliation{%
 Department of Physics and Astronomy, McMaster University, Hamilton, Ontario, L8S 4M1, Canada
}%
\affiliation{%
 Brockhouse Institute for Materials Research, Hamilton, Ontario, L8S 4M1, Canada
}%
\affiliation{%
 Canadian Institute for Advanced Research, Toronto, Ontario M5G 1M1, Canada
}%

\date{\today}

\begin{abstract}
Compelling evidence exists for electronic phase separation in cuprate high-$T_c$ superconductors, emerging near 1/8 hole doping. At these dopings and low temperatures, intertwined charge and spin stripes coexist with more uniformly doped regions in the two-dimensional (2$D$) copper-oxide planes. Each region is capable of developing superconducting pairing, either as a pair density wave (PDW) within the stripes or as a uniform $d$-wave condensate ($d$-SC) in the more homogeneous regions. Using neutron scattering on single crystals of La$_{1.875-y}$Nd$_{y}$Sr$_{0.125}$CuO$_4$, we demonstrate that the onset temperatures for spin stripe order ($T_N$) and superconductivity ($T_c$) merge as the average ordered moment vanishes in LSCO ($y = 0$), whereas Nd doping stabilizes static stripe order and suppresses $T_c$. Because the spin stripes possess the same in-plane periodicity (8$a$) as the PDW and establish the framework in which the PDW resides, the stabilization of spin stripe order enhances PDW correlations. Thus, the competition between $d$-wave pairing in the uniform regions and PDW pairing in the stripe-ordered regions can be controlled by the Nd concentration in La$_{1.875-y}$Nd$_{y}$Sr$_{0.125}$CuO$_4$, allowing the superconducting $T_c$ to vary by nearly an order of magnitude at a fixed 1/8 hole doping.
\end{abstract}

\maketitle

\section{Introduction}
\label{Sec:I}

High-temperature superconductivity (SC) in cuprates emerges from doping a Mott insulator, resulting in a complex landscape of intertwined electronic orders \cite{15_fradkin,15_keimer,19_proust,20_agterberg}. Among these orders, the pair density wave (PDW)—a spatially modulated two-dimensional superconducting state in which Cooper pairs acquire finite momentum—has become a key candidate to unify phenomena such as charge density waves (CDWs), spin stripes, and the pseudogap phase \cite{07_berg,09_berg,15_agterberg}. The observation of extreme anisotropy in the superconducting response of La$_{2-x}$Ba$_x$CuO$_4$ (LBCO) at $x = 1/8$, where 2$D$ superconductivity sets in at $T_c^{2D} \sim 40$ K but three-dimensional (3$D$) coherence is suppressed until $T_c^{3D} \sim 4$ K, provided the first experimental indication of PDW order \cite{07_li}. In LBCO, PDW order is stabilized by static spin and charge stripes that are pinned by the low-temperature tetragonal (LTT) crystal structure, thereby frustrating interlayer phase coherence and reducing the bulk 3$D$ SC transition by roughly a factor of ten relative to the 2$D$ onset \cite{21_tranquada}.

Beyond LBCO, stripe-like density waves have been observed in several cuprate families, including \ch{YBa_2Cu_3O_{7-$x$}} (YBCO) and Bi$_2$Sr$_2$Ca$_{n-1}$Cu$_n$O$_{2n+4+x}$ (BSCCO), as well as in the La-214 family. This suggests that PDW physics may represent a universal feature in the underdoped regime \cite{21_huang,21_wang,24_choi}. Numerical studies of doped Hubbard models further support this view, identifying the PDW as a competitive ground state to uniform $d$-wave superconductivity ($d$-SC), with the two orders mutually suppressing one another \cite{02_himeda,14_corboz,20_choubey,23_setty}. However, a key question remains: Can PDW correlations persist in systems such as La$_{2-x}$Sr$_x$CuO$_4$ (LSCO)—a sister 214 cuprate to LBCO but lacking LTT symmetry at low temperatures, which relegates stripe order to a more disordered regime? A promising venue to address this question is the La-214 family at 1/8 hole doping, where the superconducting $T_c$ can be anomalously suppressed by nearly a factor of ten. In these materials, spin stripe order, and by extension PDW correlations, can be probed via neutron scattering on large single crystals.

\begin{figure}[!hbp]
\includegraphics[width=\columnwidth]{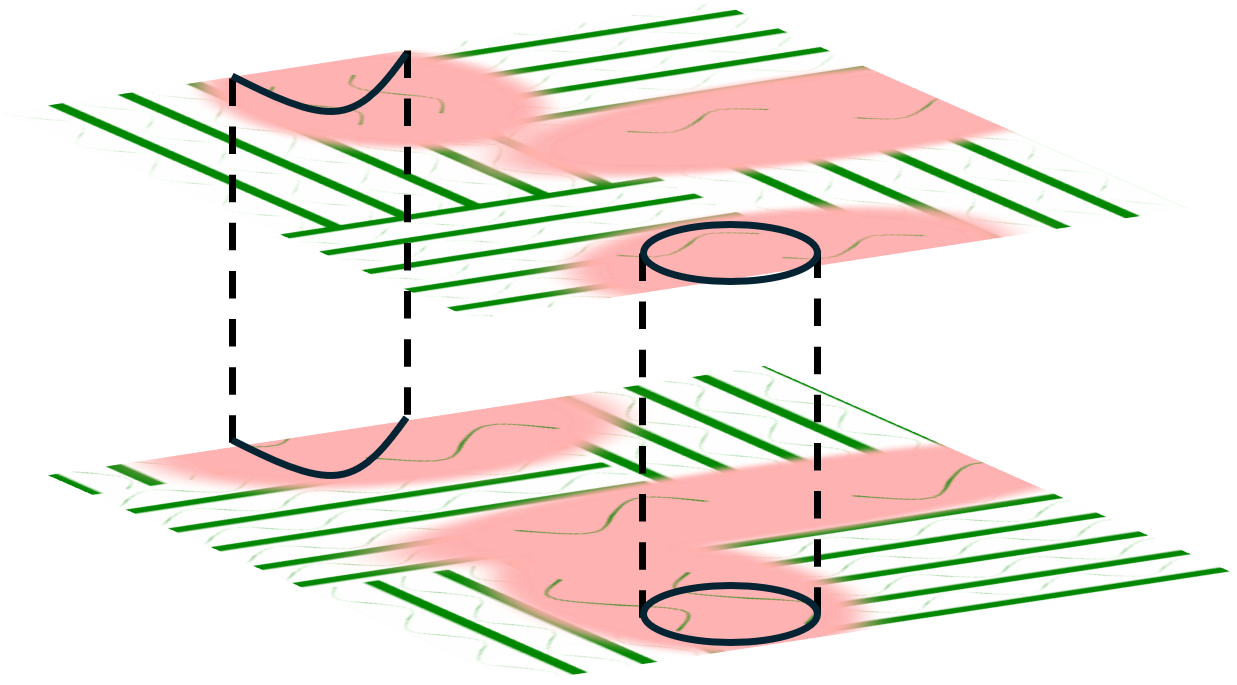}
\caption{Schematic illustration of electronic phase separation in layered 214 cuprate superconductors. The competing phases are the striped (green) regions with putative 2$D$ PDW pairing and the more uniformly doped (red) regions with uniform $d$-wave pairing. In both phases, 3$D$ superconductivity requires Josephson tunneling between layers, which is more effective in the uniform $d$-wave regions. In contrast, 3$D$ PDW superconductivity is frustrated by orthogonal stacking of the stripe (and hence PDW) structures along the $c$-axis. The schematic shown is for an approximate 50:50 volume fraction between the two phases; across the Nd-LSCO series studied here, this ratio varies from roughly 25:75 (PDW:$d$-SC) in LSCO to about 90:10 in Nd-0.4. The vertical black rods represent uniform-hole-density domains in adjacent layers, which couple via the Josephson effect to produce 3$D$ SC.
}
\label{Fig1}
\end{figure}

In this work, we propose that the 1/8 anomaly arises from an electronic phase separation between stripe regions dominated by PDW correlations and domains with uniform, hole-disordered $d$-SC. This phase separation is illustrated qualitatively in Fig. \ref{Fig1}, showing coexisting stripe regions and hole-disordered $d$-SC domains in two-dimensional sheets, with the stripe orientation rotated by 90$\degree$ between adjacent layers, a consequence of the LTT structure. Early evidence for such electronic inhomogeneity in the La-214 cuprates was the breakdown of the Yamada law \cite{98_yamada}, wherein the increase in hole doping accommodated by a stripe structure is reflected in a proportional change in the incommensurate spin and charge stripe wavevectors. However, the Yamada law holds only up to approximately 1/8 hole doping; beyond which, the periodicity saturates, implying the formation of higher hole doping uniform regions.

In LBCO, the LTT-driven stripe pinning stabilizes PDW order and suppresses 3$D$ uniform $d$-SC coherence. In contrast, the low-temperature orthorhombic (LTO) structure in LSCO permits more disordered stripe order, thereby supporting a larger volume fraction of hole-disordered $d$-SC domains that can percolate, yielding a higher-temperature 3$D$ superconducting state.

To explore this scenario, we study large single crystals of La$_{1.875-y}$Nd$_{y}$Sr$_{0.125}$CuO$_4$ ($y = 0, 0.1, 0.2, 0.4$), a tunable La-214 system bridging LSCO and Nd-LSCO. By varying the Nd concentration, we control the low-temperature structural symmetry (LTO vs. LTT) and the stripe volume fraction, thereby elucidating how the competition between PDW and $d$-SC orders governs the 1/8 anomaly. Elastic neutron scattering reveals that increasing Nd doping enhances stripe order while suppressing superconducting $T_c$, with the onset temperatures of the spin stripes ($T_N$) and superconductivity ($T_c$) converging in LSCO. This inverse relationship underscores the microscopic phase separation: PDWs within the stripes compete with uniform $d$-SC for volume fraction, with their coexistence mediated by structural pinning induced by the interplanar Nd dopants. Our results unify the La-214 family under a single framework, in which the 1/8 anomaly marks a crossover between PDW-dominated and $d$-SC-dominated regimes, thereby advancing our understanding of cuprate superconductivity by positioning PDW order as an intrinsic component of the phase diagram.

\section{Experimental Details}
\label{Sec:II}

Large single crystals of La$_{1.875-y}$Nd$_{y}$Sr$_{0.125}$CuO$_4$ ($y = 0, 0.1, 0.2, 0.4$) were grown using the traveling-solvent floating zone technique in a four-mirror image furnace. Detailed descriptions of similar growths are available elsewhere \cite{20_dragomir,22_ma2}. The resulting single crystals, cylindrical in shape, had masses ranging from 3.12 g to 6.35 g. All samples were post-annealed in oxygen at 950 $^{\circ}$C for 96 hours. For brevity, we refer to the four samples as LSCO, Nd-0.1, Nd-0.2, and Nd-0.4.

Magnetization measurements were performed on small single crystals, cut from the larger specimens, using a Quantum Design MPMS superconducting quantum interference device (SQUID) magnetometer. Magnetization measurements were performed in a small dc field of 0.5 mT applied along the crystallographic $c$-axis. Throughout the paper, we employ tetragonal notation with $a \simeq 3.8$ $\text{\AA}$ and $c \simeq 13.1$ $\text{\AA}$. 

Elastic neutron scattering measurements were conducted on the fixed-incident-energy triple-axis spectrometer VERITAS at Oak Ridge National Laboratory. In our elastic scattering studies, an incident energy of $E_i = 14.5$ meV was used. A collimation setting of 40'-40'-40'-80' provided an energy resolution of approximately 1 meV (FWHM) at the elastic position. The single crystal samples were mounted with the $(H K 0)$ plane coincident with the scattering plane.

\section{Experimental Results}
\label{Sec:III}

\subsection{Magnetic Susceptibility Measurements}
\label{Sec:IIIa}

\begin{figure}[!hbp]
\includegraphics[width=\columnwidth]{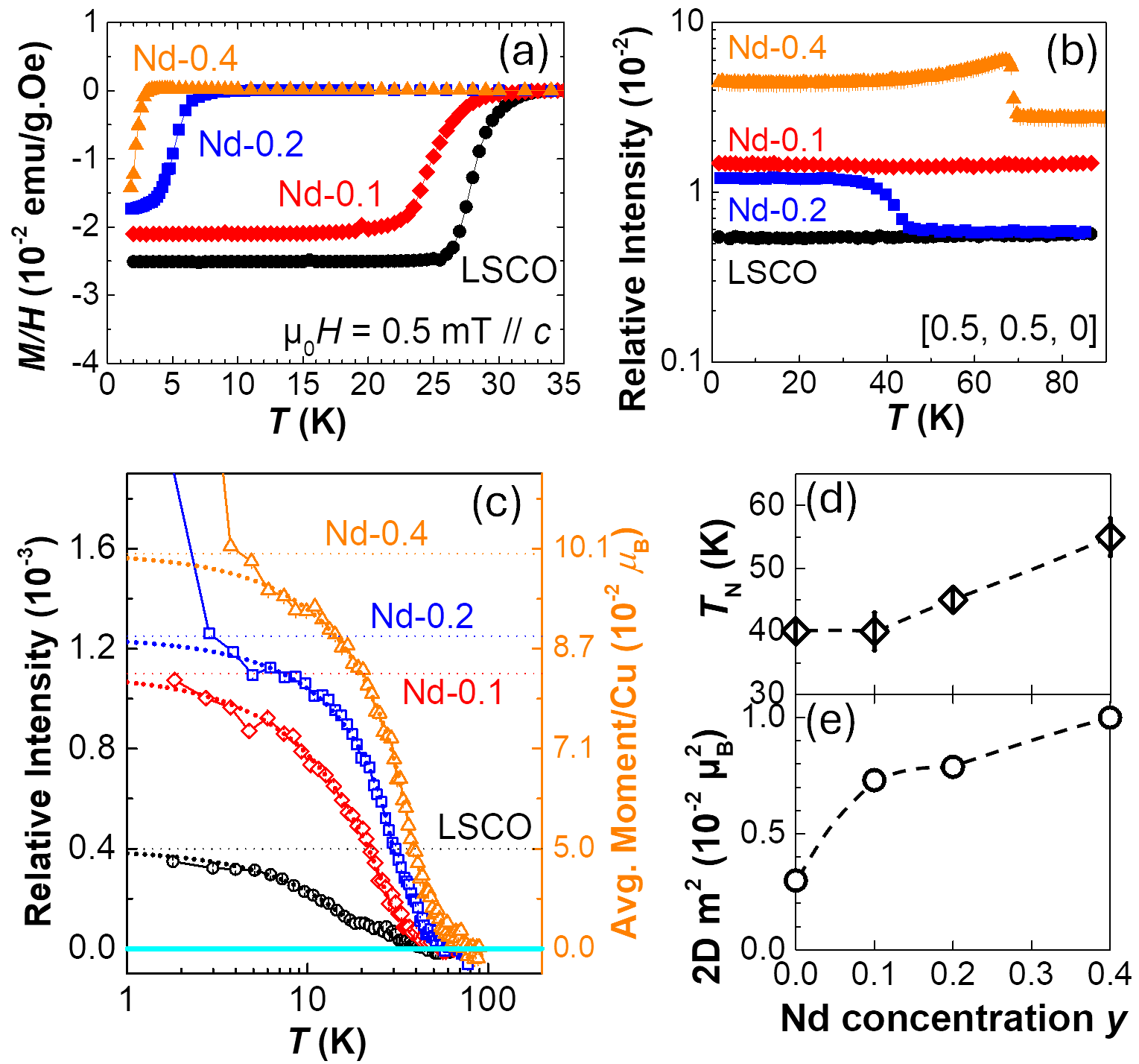}
\caption{(a) Zero-field-cooled (ZFC) magnetization measurements of the superconducting transition temperatures for all four samples. (b) The peak intensity of the [0.5, 0.5, 0] nuclear Bragg peak, normalized by the integrated intensity of the nuclear [2, 0, 0] peak, for all samples. (c) The order parameters of the incommensurate (IC) spin stripe order, with an estimated high-temperature background subtracted and normalized by the integrated intensity of the nuclear [2, 0, 0] peak. (d) The onset temperatures $T_N$ of the spin stripe order, derived from (c), for all samples. (e) The low-temperature Bragg intensity representing the squared average ordered moment associated with spin stripe order, for all samples.
}
\label{Fig2}
\end{figure}

\begin{figure}[!tbp]
\includegraphics[width=\columnwidth]{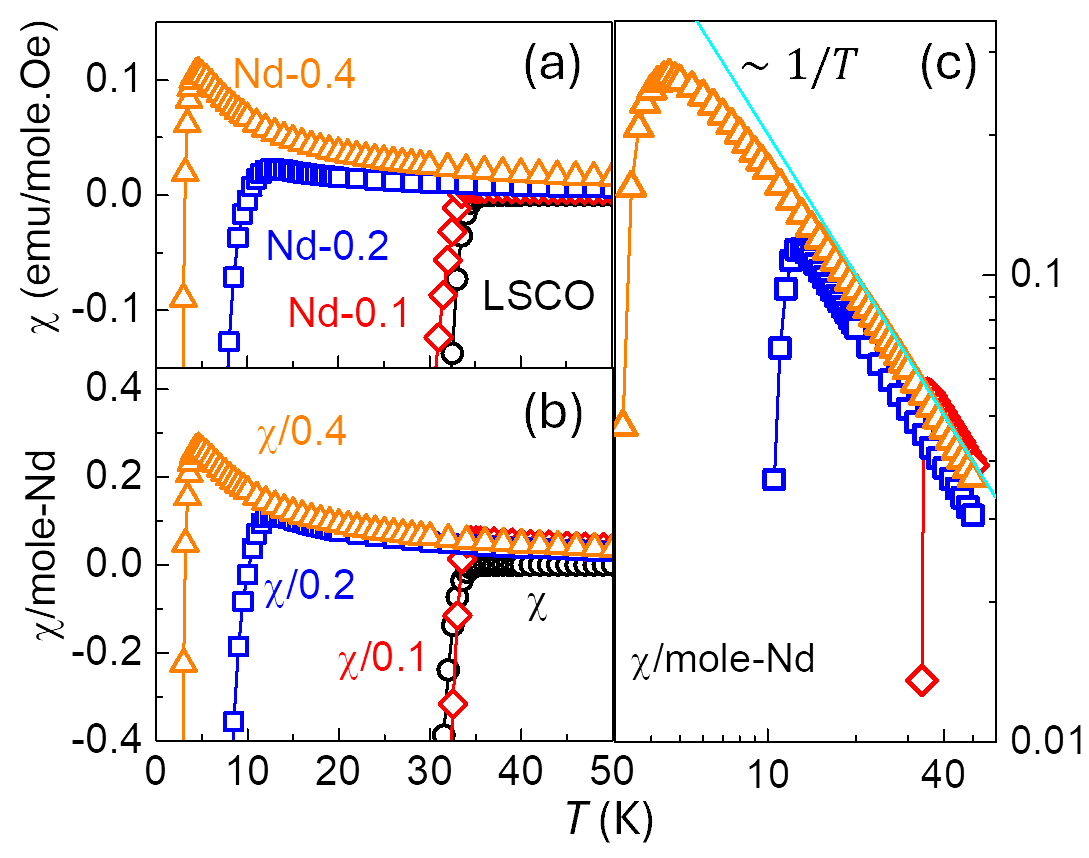}
\caption{Zero-field-cooled (ZFC) magnetization for 1/8-doped Nd-LSCO. (a) The same data as in Fig. \ref{Fig2}(a), zoomed in to show the downturn in susceptibility marking the onset of superconductivity. (b-c) The susceptibility normalized by Nd content (except for LSCO). The straight line in (c) represents the $1/T$ Curie-law behavior expected for paramagnetic Nd$^{3+}$ moments at high temperatures.
}
\label{Fig3}
\end{figure}

Figure \ref{Fig2}(a) displays the magnetic susceptibility data ($M/H$) for the four samples on a full scale, allowing a clear view of the diamagnetic signal. In general, LSCO and Nd-0.1 exhibit relatively high superconducting transition temperatures, $T_c \sim 27$ K, whereas Nd-0.2 and Nd-0.4 show significantly lower $T_c$ values of approximately 5 K or less. It is also insightful to examine the superconductivity onset marked by the downturn of a Curie-like susceptibility (notably in the Nd-doped samples), as presented in Fig. \ref{Fig3}. Figure \ref{Fig3}(a) displays the low-temperature susceptibility (below 50 K) for all samples, while Figs. \ref{Fig3}(b) and (c) show the same data normalized by the Nd content (except for LSCO).

In Fig. \ref{Fig3}(c), the susceptibility data for Nd-0.1, Nd-0.2, and Nd-0.4 are plotted on a log-log scale, so that power-law dependences appear as straight lines. A reference straight line with a slope of 1, corresponding to a $1/T$ Curie-law dependence (expected for paramagnetic Nd$^{3+}$ moments) is included. The data shows that the Curie-law behavior is observed down to approximately 10 K. This behavior is expected since the diluted Nd$^{3+}$ moments between the copper-oxide planes are only weakly coupled to either the Cu$^{2+}$ moments or to one another and therefore remain largely paramagnetic. The contribution from the Cu$^{2+}$ moments is minimal due to their small size and the strong antiferromagnetic interactions that effectively cancel the net paramagnetic response.

An onset temperature for superconductivity, $T_c^{(onset)}$, is readily identified from the sharp peak in the susceptibility versus temperature plots, while the midpoint, $T_c^{(mid)}$, is determined from Fig. \ref{Fig2}(a). In all cases, $T_c^{(onset)}$ is significantly above $T_c^{(mid)}$, as summarized in Table \ref{Tab1}.

\subsection{Elastic Neutron Scattering Measurements}
\label{Sec:IIIb}

Elastic neutron Bragg intensity measurements at the commensurate peak [0.5, 0.5, 0] as a function of temperature were carried out for all four single crystal samples to probe the LTO-to-LTT structural phase transition. Figure \ref{Fig2}(b) shows the [0.5, 0.5, 0] Bragg intensity normalized to the integrated intensity of the nuclear [2, 0, 0] peak for each sample. Structural transitions are observed at approximately 68 K and 40 K in Nd-0.4 and Nd-0.2, respectively, while no such transitions are detected in LSCO and Nd-0.1 down to 1.5 K. The LTO-LTT structural transition has also been observed at 68 K in Nd-0.4 single crystals in splittings of the [3, 3, 0] x-ray Bragg reflections \cite{20_dragomir}. Earlier work by Crawford \textit{et al.} reported similar superconducting $T_c$ and structural transition temperatures $T_S$ in polycrystalline La$_{1.88-y}$Nd$_{y}$Sr$_{0.12}$CuO$_4$ ($0 \le y \le 0.6$) \cite{91_crawford}, and our single crystal measurements are consistent with these findings. Similarly, Tajima \textit{et al.} found an LTO-LTT phase boundary near $y \approx 0.12$ for Nd-LSCO at $x = 0.15$ \cite{01_tajima}, in close agreement with our results.

\begin{table*}[t]
\caption{Summary of results for 1/8 hole-doped La-214 cuprates. The notation $n$ and $\mu$ for $T_{SO}$ indicates measurements by neutron scattering and $\mu$SR, respectively, as described in the text.}
\centering
\begin{tabular}{c|cc|c|c|c|c|cc}
\hline\hline
Compound at & \multicolumn{2}{c|}{Superconducting $T_c$} & Onset of  & Onset of 2D SC & Onset of PDW & LTO-LTT &  \multicolumn{2}{c}{Average Cu moment $\mu_\text{av}$}  \\
1/8 hole-doping & $T_c^{(onset)}$ & $T_c^{(mid)}$  & SDW order $T_N$ & $T_c^{2D}$ & $T_{SO}$ & $T_S$ & $T=0$ & $T=T_c^{(mid)}$ \\[1ex]
\hline
LBCO    &  40 K & 4 K  & 50 K & 38 K & 38 K ($n,\mu$) & 55 K        & 0.1 $\mu_B$    & 0.1 $\mu_B$ \\
\hline
Nd-0.4  & 4.6 K & 2.5 K& 55 K & 25 K & 33 K ($n, \mu$) & 68 K & 0.1 $\mu_B$    & 0.1 $\mu_B$ \\
Nd-0.2  & 13 K  & 5 K  & 45 K & -    & 25 K ($n$) & 40 K     & 0.089 $\mu_B$  & 0.083 $\mu_B$ \\
Nd-0.1  &  35 K & 25 K & 40 K & -    & -    & $< 1.5$ K   & 0.083 $\mu_B$  & 0.042 $\mu_B$ \\
LSCO    &  37 K & 28 K & 40 K & -    & 17 K ($n, \mu$) & $< 1.5$ K   &  0.050 $\mu_B$ & 0.019 $\mu_B$  \\
\hline
Fe-LSCO &  8 K  & 4 K  & 40 K & 21 K \cite{21_huang} & 17 K ($n, \mu$ \cite{12_suzuki}) & $< 10$ K   & $\sim$0.07 $\mu_B$ & -\\[1ex]
\hline\hline
\end{tabular}
\label{Tab1}
\end{table*}

Figure \ref{Fig2}(c) presents the temperature dependence of the parallel spin stripe order parameter, measured by elastic neutron scattering for Nd-0.4, Nd-0.2, Nd-0.1, and LSCO, after subtracting a fitted high-temperature background. For LSCO and Nd-0.1, the incommensurate (IC) stripe peaks are shifted by approximately 2–3$\degree$ relative to the Cu–O bond direction in the CuO$_2$ plane, as evidenced by the slight displacement of the peaks from the nominal [$0.5 \pm \delta$, 0.5, 0] and [0.5, $0.5 \pm \delta$, 0] positions in reciprocal space (with $\delta \approx 0.125$). Thus, these peaks are observed at [$0.5 \pm \delta$, $0.5 \mp \epsilon$, 0] and [$0.5 \mp \epsilon$, $0.5 \pm \delta$, 0] (with $\epsilon \approx 0.006$). This slanted stripe order was first identified in La$_2$CuO$_{4+y}$ \cite{99_Lee} and has since been discussed in LSCO and related compounds \cite{00_kimura,00_Katano,02_Fujita,24_he,22_Wang}.

To compare quantitatively the spin stripe order parameters in the four crystals, the observed elastic intensities of the IC peaks were normalized to the integrated intensity of the [2, 0, 0] nuclear Bragg peak. As shown in Fig. \ref{Fig2}(c), both Nd-0.4 and Nd-0.2 exhibit a marked enhancement in the order parameter below approximately 4 K. This enhancement is attributed to the coupling between static Nd$^{3+}$ and Cu$^{2+}$ magnetic moments, which promotes the development of 3$D$ spin stripe correlations at low temperatures \cite{96_tranquada}. A similar effect is weak in Nd-0.1 and absent in LSCO.

A major finding of this study, illustrated in Fig. \ref{Fig2}(c), is that the parallel spin stripe order increases monotonically with Nd doping. Assuming a common spin stripe structure and volume fraction for all samples (with the magnetic Bragg intensity proportional to the square of the ordered moment) and adopting the previously estimated ordered Cu moment of approximately 0.1 $\mu_B$ in Nd-0.4 \cite{96_tranquada}, we deduce that the \textit{average} ordered Cu moment at zero temperature is roughly 0.05 $\mu_B$ in LSCO, 0.083 $\mu_B$ in Nd-0.1, and 0.089 $\mu_B$ in Nd-0.2. These values are summarized in Table \ref{Tab1} and plotted in Fig. \ref{Fig2}(e). This analysis neglects the low-temperature enhancement of the order parameter in Nd-0.2 and Nd-0.4 due to Nd$^{3+}$–Cu$^{2+}$ coupling. An earlier elastic neutron scattering study comparing the average Cu moments in LSCO and Nd-0.4 at 1/8 doping reported a ratio of 0.33(8) \cite{08_changj}, which is reasonably close to our current result of 0.5. The deduced Cu moment of around 0.05 $\mu_B$ in our LSCO sample is consistent with a previously reported value of 0.07(1) $\mu_B$ \cite{24_he} and is approximately half the value observed in oxygen-doped La$_2$CuO$_{4+y}$ \cite{99_Lee}.

We use the term ``averaged moment" because neutron scattering measures an integrated response over the sample volume and does not resolve microscopic spatial variations. In contrast, muon spin relaxation ($\mu$SR) is a local probe capable of distinguishing between magnetic and non-magnetic regions in heterogeneous samples. $\mu$SR measurements on 1/8-doped LSCO and Nd-0.4 have revealed evidence for heterogeneous spin and superconducting (SC) order. Specifically, the magnetically ordered volume fraction is estimated to be low in LSCO (20\% \cite{02_savici} and 40\% \cite{05_savici}), in contrast to nearly 100\% in Nd-0.4 \cite{17_guguchia}. $\mu$SR measurements can also probe the superconducting volume fraction, which is estimated to be about 50\% \cite{05_savici} in LSCO. A trade-off between superconductivity and the magnetic volume fraction has been reported in La$_{2-x-y}$Eu$_y$Sr$_x$CuO$_4$ (LESCO), suggesting that SC and static spin order reside in microscopically separated regions \cite{03_kojima}. Our neutron diffraction results, showing an increasing average moment with Nd concentration, are consistent with these observations.

The ``averaged" Cu moment increased with increasing Nd doping can be explained by (1) enhanced Cu moment upon doping, or (2) increased domain population upon doing. Previous $\mu$SR \cite{02_savici,05_savici,17_guguchia} studies rule out the first scenario. We therefore assume that the parallel spin stripe structure and the ordered moment are roughly uniform across our LSCO and Nd-LSCO series at $x = 1/8$, but that the stripe phase volume fraction increases in proportion to the average ordered Cu$^{2+}$ moment. Assuming our Nd-0.4 sample is approximately 90\% occupied by the PDW phase at low temperatures, the corresponding PDW occupation in LSCO can be estimated from the ratio of the normalized low-temperature magnetic Bragg intensities between LSCO and Nd-0.4, yielding a PDW occupation of roughly 20--25\%, in agreement with $\mu$SR estimates.

\begin{figure}[!htbp]
\includegraphics[width=\columnwidth]{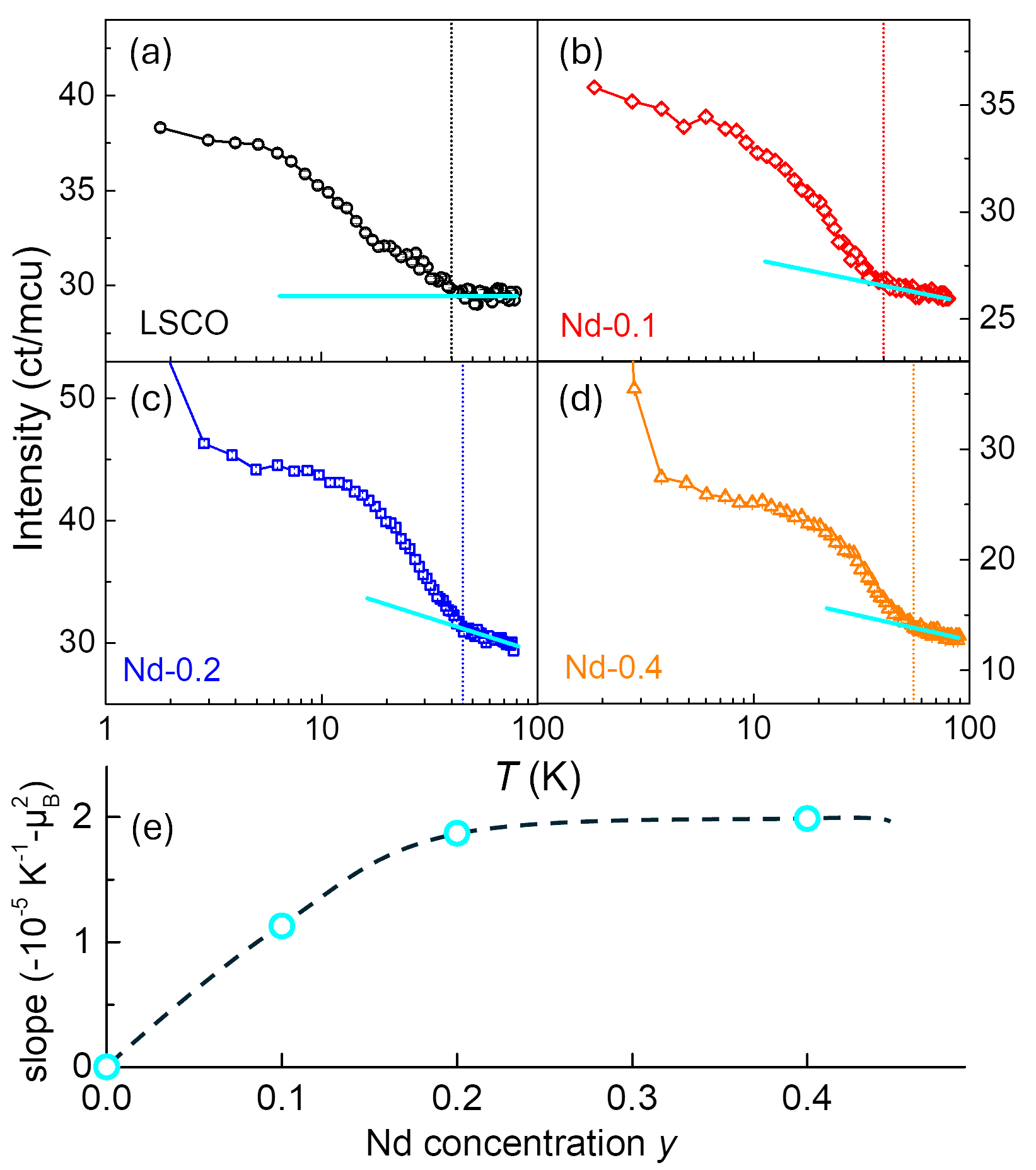}
\caption{(a--d) Spin stripe order parameters versus temperature plotted on a logarithmic temperature scale. The vertical dashed lines indicate $T_N$. The light blue lines represent fits to the high-temperature background. In (e) the background slope for each of the LSCO and Nd-LSCO samples is summarized, showing a linear increase from zero in LSCO that saturates for Nd doping of 0.2 and beyond.
}
\label{Fig4}
\end{figure}

Closer inspection of the order parameter near $T_N$ (Fig. \ref{Fig4}(a)--(d)) reveals that the high-temperature background is temperature-dependent, decreasing with increasing temperature in all Nd-containing samples. This observation implies that in Nd-0.1, Nd-0.2, and Nd-0.4, short-range incommensurate spin stripe correlations persist up to approximately 100 K, whereas in LSCO the correlations emerge only at $T_N = 40$ K. A robust determination of $T_N$ for Nd-0.1, Nd-0.2, and Nd-0.4 is made by identifying the temperature at which the order parameter deviates from the sloping high-temperature background. The determined $T_N$ values, indicated by vertical lines in Figs. \ref{Fig4}(a)--(d), are summarized in Fig. \ref{Fig2}(d) and listed in Table \ref{Tab1}. While $T_N$ is 40 K in LSCO and Nd-0.1, it increases to 45 K and 55 K in Nd-0.2 and Nd-0.4, respectively—a relatively modest increase.

The temperature-dependent sloping background can be quantified by correlating the relative low-temperature averaged Cu$^{2+}$ moments measured for the LSCO and Nd-LSCO samples. A given relative low-temperature spin stripe Bragg peak intensity translates to an averaged ordered moment. Figure \ref{Fig4}(e) shows that the slope of the spin stripe order parameter increases smoothly from zero for LSCO and saturates beyond approximately $y = 0.2$. Since this sloping background persists up to 90 K, it is likely present at least through the pseudogap crossover temperature, estimated to be around $T^*$ = 150 K at 1/8 doping \cite{17_doiron}. A recent nuclear magnetic resonance (NMR) work investigated the relationship between spin-stripe order and the pseudogap phase in the cuprate material La$_{1.8-x}$Eu$_{0.2}$Sr$_{x}$CuO$_4$ and showed that spin-stripes with short correlation times permeate much or all of the pseudogap regime\cite{25_missiaen}.  This is consistent with our interpretation of the observed high-temperature sloping background in Nd-LSCO with $x$ = 1/8. Both of these results appear to align well with a recent theoretical work on the Hubbard model \cite{24_simkovic}. Also, the fluctuating PDW (FPDW) has been considered as the ``mother state" that onsets at high temperatures around the pseudogap transition $T^*$ and evolves into static CDW and SDW orders at lower temperatures or under pinning conditions \cite{14_lee,18_dai,19_lee}.

One may then ask: why are these weak spin stripe fluctuations observable in Nd-LSCO but not in LSCO? Although one might naturally attribute this to the LTT crystal structure at low temperatures, Nd-0.1, like LSCO, shows no clear evidence for an LTO-LTT transition above 1.5 K, while Nd-0.2 exhibits $T_N$ slightly above $T_S$. This trend—of increasing sloping spin stripe background with Nd content—suggests that the background originates from short-range spin stripe and PDW fluctuations within the volume fraction that eventually develops long-range spin stripe order. These fluctuations are difficult to observe in LSCO because its stripe volume fraction is much smaller than that of the uniformly doped regions.

\section{\texorpdfstring{Two-Step Evolution of the Spin Stripe Order Parameter in LSCO $x = 1/8$}{Two-Step Evolution of the Spin Stripe Order Parameter in LSCO x = 1/8}}
\label{Sec:IIIc}

\begin{figure}[!tbp]
\includegraphics[width=\columnwidth]{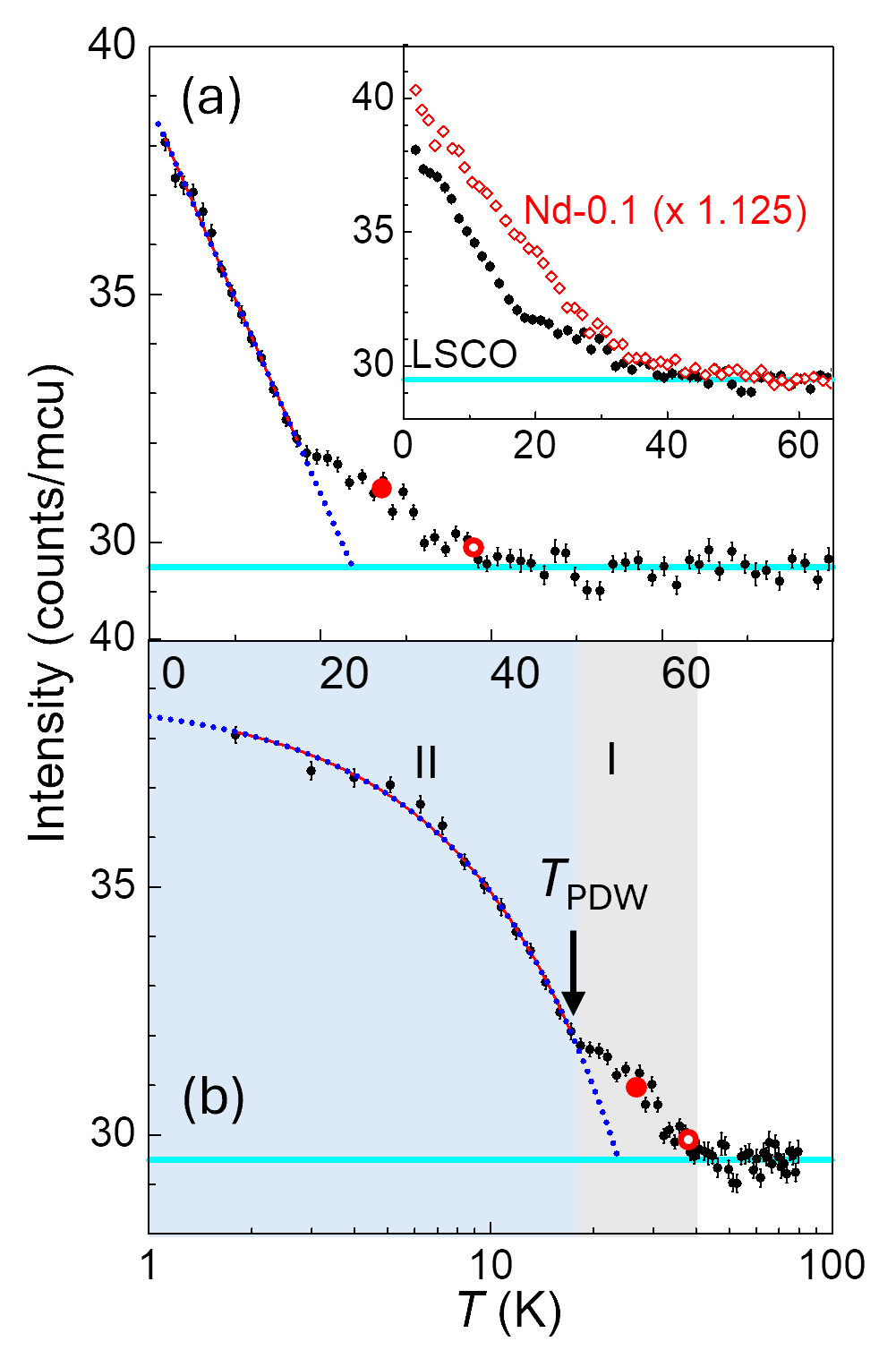}
\caption{(a--b) Two-step evolution of the spin stripe order in LSCO. (a) The order parameter of the incommensurate magnetic peak associated with parallel spin stripe order is plotted on a linear temperature scale. (b) The same data are shown on a linear-log scale. The $T_c^{(mid)}$ and $T_c^{(onset)}$ values determined from ZFC magnetization are indicated by filled and open red circles, respectively. The dashed blue line represents a linear fit to the order parameter below 17 K. The inset in (a) compares the order parameters of LSCO and Nd-0.1 (scaled for clarity).
}
\label{Fig5}
\end{figure}

A striking feature of the spin stripe order parameter in LSCO is its distinct temperature evolution compared to the Nd-LSCO samples. This is evident in Figs. \ref{Fig4}(a--d) and is more clearly shown in Figs. \ref{Fig5}(a) and (b), where the LSCO order parameter data are plotted on both linear-linear and linear-log scales. A clear kink appears in the order parameter at around 17 K, which coincides with $T_{SO}$, the temperature at which zero longitudinal field $\mu$SR measurements indicate  the onset of static magnetism on the muon time scale \cite{94_kumagai,02_savici}, and well below $T_N$ = 40 K. The spin stripe order in LSCO develops in two steps: an initial phase begins at $T_N$ = 40 K, which almost coincides with $T_c^{(onset)}$.  This saturates near the superconducting $T_c^{(mid)}$ = 28 K, and is followed by a second regime of approximately linear growth below $T_{SO} = T_\text{PDW} \sim 17$ K. We associate $T_\text{PDW}$ with $T_{SO}$ as they are the same in LBCO at $x$ = 1/8, and there is no other relevant temperature scale for the spin stripe order below $T_N$.  The inset in Fig. \ref{Fig5}(a) compares LSCO with Nd-0.1 (with the Nd-0.1 data scaled to align the background), highlighting the distinct behavior near $T_\text{PDW}$ even in the sample with the lowest Nd concentration and the same low temperature structure as LSCO.

In Fig. \ref{Fig5}, the onset of superconductivity in LSCO, $T_c^{(onset)}$, is marked by an open red circle, while the midpoint, $T_c^{(mid)}$, is denoted by a solid red circle. Clearly, $T_c^{(onset)}$ and $T_N$ are nearly coincident in LSCO.

The two-step behavior observed in the LSCO order parameter is consistent with earlier inelastic neutron studies, which found that the magnetic spectral weight first appears as gapless around 40 K and later develops a gap below $T_c^{(mid)} \approx 28$ K, in conjunction with the onset of superconductivity \cite{13_romer}. Moreover, recent synchrotron X-ray scattering measurements on LSCO near the 1/8 anomaly ($x = 0.115$) report short-range CDW order emerging above 70 K, with more extended order developing below $T_c$ \cite{23_wen}, mirroring the two-step evolution observed here in the spin stripe order of LSCO.

\subsection{In-Plane Spin Stripe Correlation Lengths}
\label{Sec:IIId}

\begin{figure*}[!tbp]
\includegraphics[width=\textwidth]{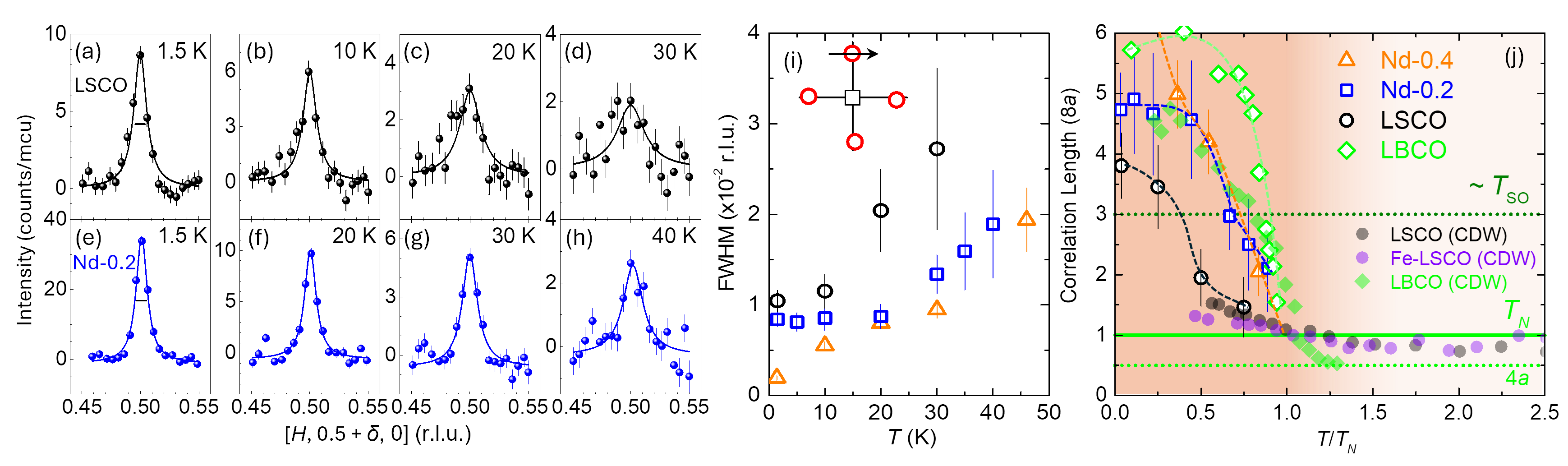}
\caption{$Q$-scans of the spin stripe order magnetic peak at various temperatures for (a--d) LSCO and (e--h) Nd-0.2. (i) Full Width at Half Maximum (FWHM) extracted from the $Q$-scans for LSCO and Nd-0.2 as a function of temperature. Data for Nd-0.4, taken from Ref. \cite{24_chen}, are also shown. An approximate resolution convolution of the line shapes, assuming the Nd-0.4 width at 1.5 K is resolution limited, is applied to estimate the intrinsic FWHM. The inset in (i) shows the quartet of IC spin stripe peaks (red circles) surrounding the commensurate [0.5, 0.5, 0] peak (black square), with the arrow indicating the $Q$-scan trajectory. (j) The corresponding in-plane correlation lengths (2/FWHM) in units of $\lambda_\text{PDW} = 8a$, plotted as a function of normalized temperature $T/T_N$. The correlation lengths for LBCO (CDW and SDW), LSCO (CDW), and Fe-LSCO (CDW) are taken from Refs. \cite{08_tranquada}, \cite{19_wen}, and \cite{21_huang}, respectively.
}
\label{Fig6}
\end{figure*}

In addition to the elastic Bragg scattering measurements, reciprocal-space scans were performed to determine the in-plane spin stripe correlation length as a function of temperature. The line scans in reciprocal space through incommensurate ordering wavevectors (e.g., [$H$, $0.5 \pm \delta$, 0]) are illustrated in the inset of Fig. \ref{Fig6}(i). Line scans for LSCO covering temperatures from 1.5 K to 30 K are shown in Figs. \ref{Fig6}(a)--(d), while scans for the Nd-0.2 single crystal from 1.5 K to 40 K are presented in Figs. \ref{Fig6}(e)--(h).

All scans were fitted with Lorentzian line shapes, yielding the measured FWHM and thus the in-plane spin correlation length ($\xi = 2/\mathrm{FWHM}$). The fits of the LSCO and Nd-0.2 data are shown as solid lines in Figs. \ref{Fig6}(a)--(h). Previously published FWHM measurements on the same Nd-0.4 crystal \cite{24_chen} were also incorporated into this comparative study.

Figure \ref{Fig6}(i) plots the FWHM as a function of temperature for LSCO, Nd-0.2, and Nd-0.4. At all temperatures, LSCO exhibits a larger FWHM than both Nd-0.2 and Nd-0.4, although for LSCO the FWHM saturates below 20 K, while that for Nd-0.2 saturates below 30 K. For Nd-0.4 the FWHM does not reach saturation even at the lowest measured temperature, presumably due to the development of 3$D$ spin stripe correlations induced by weak coupling between Nd and Cu magnetism at low temperatures. The base temperature FWHM in Nd-0.4 is considerably smaller than in LSCO or Nd-0.2, and it is assumed to be resolution limited in the 3$D$ stripe state at 1.5 K.

\section{Discussion}
\label{Sec:IV}

We begin our discussion with an analysis of the correlation lengths in these 1/8-doped crystals. Figure \ref{Fig6}(j) compiles the in-plane correlation lengths extracted from the line scan data for spin stripes in LSCO, Nd-0.2, and Nd-0.4. In addition, we include in-plane SDW and CDW correlation lengths from previously published work on LBCO at $x=1/8$ \cite{08_tranquada}, as well as in-plane CDW correlation lengths in LSCO \cite{19_wen} and 1\% Fe-doped LSCO \cite{21_huang} at 1/8 hole doping. The in-plane SDW correlation lengths for LSCO, Nd-0.2, and Nd-0.4 are derived from the FWHM data in Fig. \ref{Fig6}(i) by taking the Nd-0.4 FWHM at $T = 1.5$ K as a measure of the instrumental resolution, and then using the convolution relation $w_{o}^2 = w_{i}^2 + w_{inst}^2$, where $w_{o}$, $w_{i}$, and $w_{inst}$ denote the observed, intrinsic, and instrumental FWHM, respectively.

In Fig. \ref{Fig6}(j) these in-plane correlation lengths are plotted in units of the PDW (and SDW) periodicity, against the temperature normalized to $T_N$.  There are several interesting conclusions that can be drawn from this plot.  First, while the PDW and SDW wavelengths are 8$a$ (where $a$ is the Cu-Cu distance, $\sim$ 3.78 $\text{\AA}$), the CDW or charge stripe periodicity is half of this, 4$a$.  Given that a well-defined SDW or CDW diffraction peak will only be observed if the relevant correlation length is greater than the density wave periodicity, one expects to observe incommensurate peaks from charge stripes at higher temperatures than from spin stripes, as their periodicity is only half as large, and there will be some temperature regime where 4$a$ $\le$ $\xi$($T$) $\le$ 8$a$.

Notably, charge stripes are indeed observed at higher temperatures than spin stripes, well above $T_N$ in all of LSCO, Fe-LSCO and LBCO, all at $x$ = 1/8, and up to 2.5 $\times$ $T_N$ in LSCO and Fe-LSCO.  In addition $T_N$ itself is coincident with both the CDW and SDW correlation lengths crossing $\lambda_\text{PDW}$ = 8$a$ ($\sim$ 30 $\text{\AA}$) for all the 1/8 doped La-214 cuprate samples considered.  Finally, the spin ordering temperature associated with zero field $\mu$SR measurements, $T_{SO}$, is lower than $T_N$ in all cases and corresponds roughly to the temperature at which the SDW correlation length crosses 3 $\times$ $\lambda_\text{PDW}$ ($\sim$ 90 $\text{\AA}$).

The PDW state, proposed as a distinct superconducting state in LBCO at $x=1/8$, is supported by experimental signatures such as the strong 2$D$ superconductivity emerging at $T_c^{2D} \approx 40$ K \cite{07_li}, which coincides with the static spin order temperature ($T_{SO}$) observed by both $\mu$SR and neutron scattering \cite{91_luke,08_tranquada}, while the 3$D$ superconducting order only emerges at much lower temperatures ($T_c^{(mid)} \approx 4$ K). At $T_c^{2D}$, the in-plane resistivity $\rho_{ab}$ drops by an order of magnitude or more, whereas the out-of-plane resistivity $\rho_c$ remains high. This pronounced anisotropy suggests that while the charge stripes become superconducting, their rotation by 90$\degree$ between layers (consistent with a PDW state), frustrates interlayer Josephson coupling \cite{07_berg,09_berg}. Consequently, although PDW order enables 2$D$ superconductivity, it suppresses 3$D$ coherence. At temperatures below approximately 5 K, 3$D$ superconductivity emerges in a filamentary manner, likely associated with isolated regions of more-uniform hole concentration that favor uniform $d$-wave SC \cite{21_tranquada,22_lozano,23_ren}. In this scenario, percolative 3$D$ coherence is achieved only at low temperatures, when these spatially separated, uniform $d$-SC domains couple through the $c$-axis, as schematically illustrated in Fig. \ref{Fig1}. Furthermore, the uniform $d$-wave regions are expected to compete with PDW order, reinforcing the picture of microscopic phase separation where filamentary 3$D$ SC and spin-stripe order originate from distinct regions.

\begin{figure}[!tbp]
\includegraphics[width=\columnwidth]{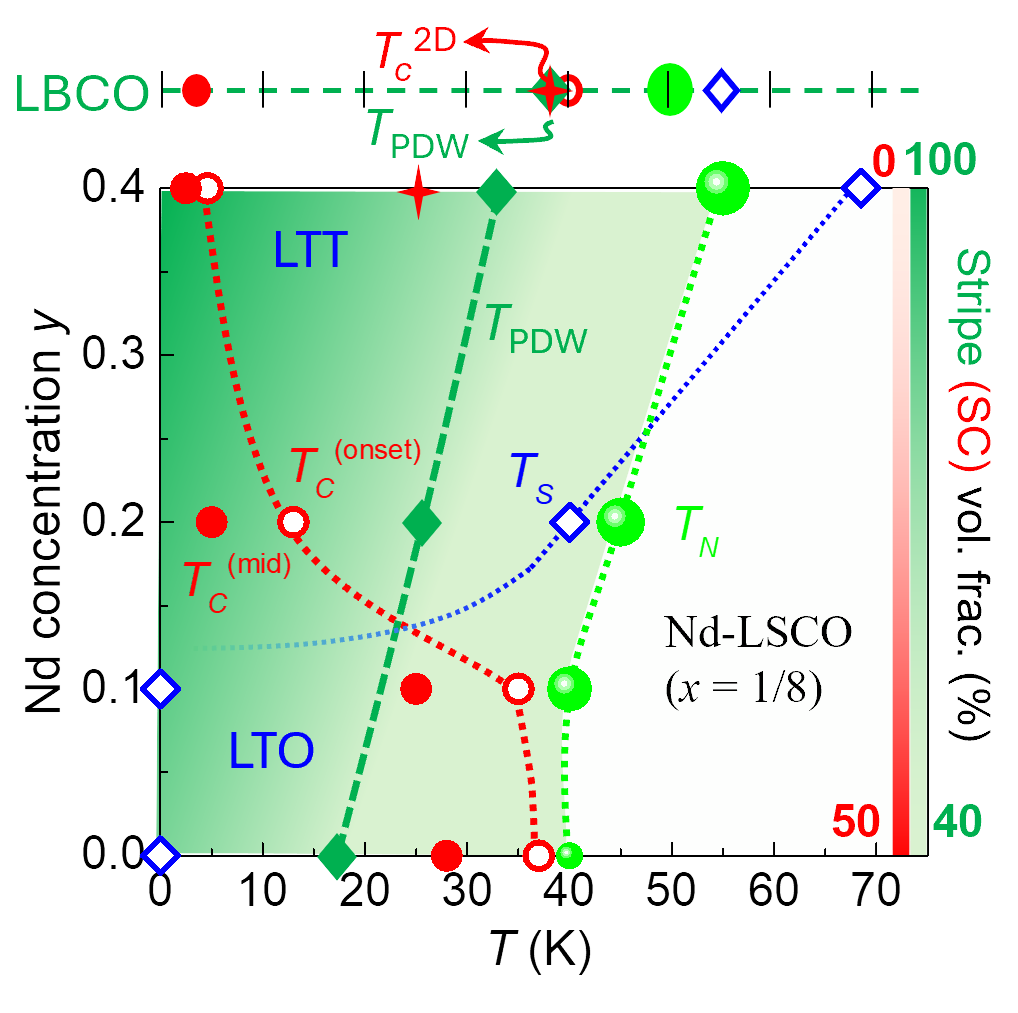}
\caption{Nd-content versus temperature phase diagram for La$_{1.875-y}$Nd$_{y}$Sr$_{0.125}$CuO$_4$, showing the measured values of $T_S$ (open diamond), $T_c^{(mid)}$ (filled circle), $T_c^{(onset)}$ (open circle), $T_N$ (green ball), $T_c^{2D}$ (red cross), and $T_{SO}$ (filled diamond) as listed in Table \ref{Tab1}. The regions corresponding to the PDW state are shaded in green. LBCO data are also included where a putative PDW state has been observed. Note that the LTO-to-LTT structural phase boundary appears between $0.1 < y < 0.2$. The size of the $T_N$ symbols is proportional to the low-temperature ordered moment in the spin stripe structure. Stripe and SC volume fractions, as measured by $\mu$SR, are indicated for LSCO and Nd-0.4.
}
\label{Fig7}
\end{figure}

Besides LBCO ($x=1/8$), signatures of 2$D$ superconductivity—a hallmark of PDW order—have been observed in Fe-LSCO \cite{21_huang} and in single-layered cuprates such as Nd-LSCO and LESCO \cite{08_ding,20_shi_NC,20_shi_SA}. In these cases, strong 2$D$ superconducting correlations are marked by a sharp reduction in the in-plane resistivity $\rho_{ab}$. For LBCO and Nd-0.4 at $x=1/8$, $T_c^{2D}$ is indicated by red crosses in Fig. \ref{Fig7}, which compiles our measurements for the La$_{1.875-y}$Nd$_{y}$Sr$_{0.125}$CuO$_4$ series along with published data for LBCO at $x=1/8$. Fig. \ref{Fig7} also depicts the $T_c^{(mid)}$ and $T_c^{(onset)}$ values obtained from ZFC magnetization (depicted as filled and empty red circles, respectively). In some cases, $T_c^{2D}$ exceeds $T_c^{(onset)}$ (they are nearly equal in LBCO), but they are always closely aligned with $T_{SO}$, suggesting that $T_c^{2D}$ and $T_{SO}$ measure two facets (electronic and magnetic) of the same phase - the PDW state. 

For LBCO, which lacks magnetic dopants, $T_c^{(onset)}$ essentially coincides with $T_c^{2D}$. In Nd- and Fe-doped LSCO as well as in LBCO, 3$D$ superconductivity is achieved below $T_{SO}$. However, in LSCO, $T_c^{(mid)} = 28$ K exceeds $T_{SO} = 17$ K, implying that the 3$D$ SC in LSCO arises primarily from uniform $d$-wave SC in the hole-disordered regions (see Fig. \ref{Fig1}). In Nd-containing samples, the presence of Nd and the LTT distortion diminishes the volume of the hole-disordered region, leading to enhanced stripe order, as evidenced by both $\mu$SR measurements and our elastic neutron scattering results. This is consistent with the measured in-plane spin stripe correlation lengths being considerably smaller in LSCO than in Nd-0.2 and Nd-0.4 at all temperatures, as shown in Fig. \ref{Fig6}, and can be attributed to the electronic phase separation and the small volume fraction occupied by the spin stripe/PDW phase in LSCO at $x$ = 1/8.   

Moreover, as just described, the spin ordering temperature $T_{SO}$ associated with zero field $\mu$SR is always lower than the $T_N$ defined by the onset of the spin stripe Bragg peaks determined from neutron scattering. For the Nd-LSCO at $x$ = 1/8 series presented here, the spin stripe order parameter shows only an inflection point at $T_{SO}$ \cite{24_chen}, the temperature dependence of the in-plane correlation length appears to capture $T_{SO}$, as the correlation length saturates below $T_{SO}$. In LSCO, the elastic order parameter exhibits an abrupt change at $T_{SO}$, coinciding with the onset of static magnetism as seen in zero-field $\mu$SR. We associate this clear anomaly in the LSCO spin stripe order parameter with the formation of a PDW, which requires a spin stripe correlation length well beyond $\lambda_\text{PDW}$ = 8$a$, and we will henceforth refer to $T_{SO}$ as $T_\text{PDW}$. Of course, this begs the question of why the formation of the PDW has such a clear signature in elastic neutron scattering in LSCO, but not for Nd-LSCO or LBCO.  It is at least plausible that this is due to the relative lack of disorder in LSCO compared with LBCO at the same doping, itself a result of the relative ionic size mismatch of Sr$^{2+}$ compared to Ba$^{2+}$ and the La$^{3+}$ that both of these dopants replace, as well as the fact that LSCO contains no Nd$^{3+}$ impurities.

Figure \ref{Fig7} summarizes the phase diagram for the La-214 cuprates at $x=1/8$, compiling data on the LTO-LTT structural transition ($T_S$), the onset of spin stripe order ($T_N$), the PDW onset ($T_{SO}$), and the superconducting transition temperatures from ZFC magnetization ($T_c^{(onset)}$ and $T_c^{(mid)}$), along with the 2$D$ SC transition ($T_c^{2D}$) where available. The size of the green circles (representing $T_N$) is proportional to the averaged low-temperature ordered Cu$^{2+}$ moment. Assuming that the size of the ordered moment within a stripe is solely determined by the hole doping ($x = 1/8$), the measured averaged Cu moment serves as a proxy for the relative volume fraction of the PDW phase. Consequently, LSCO at $x = 1/8$ exhibits a smaller PDW volume fraction, while Nd-0.4 and LBCO at $x = 1/8$ show larger PDW volume fractions, in agreement with previous $\mu$SR observations.

Finally, Fig. \ref{Fig7} clearly demonstrates that the phase diagram of these La-214 cuprates at $x = 1/8$ is divided into two distinct ground states by the crystallographic structure. Nd-0.4, Nd-0.2, and LBCO all exhibit an LTO-LTT structural transition near or above $T_N$, whereas LSCO and Nd-0.1 remain in the LTO structure down to 1.5 K. As argued above, Nd-0.4, Nd-0.2, and LBCO possess a high PDW volume fraction (approaching 100\%), with most superconducting pairing occurring within the PDW phase, resulting in 2$D$ PDW superconductivity that sets in at $T_{SO}$ while 3$D$ SC is suppressed to very low temperatures (around 5 K) due to the 90$\degree$ rotation of the PDW between layers. In contrast, for LSCO and Nd-0.1, $d$-SC pairing occurs within the relatively large uniform hole disordered volume fraction and this pairing can achieve 3$D$ coherence at relatively high temperatures by Josephson coupling to other uniform hole-disordered regions in neighboring planes, as illustrated qualitatively in Fig. \ref{Fig1}.  Hence 3$D$ SC in LSCO and Nd-0.1 at $x$ = 1/8 occurs first within the uniform regions, and is already well developed above $T_\text{PDW}$.

\section{Conclusions}
\label{Sec:V}

We have performed magnetic susceptibility and elastic neutron scattering studies on four single crystals of La$_{1.875-y}$Nd$_{y}$Sr$_{0.125}$CuO$_4$. Combined with earlier measurements on La$_{1.875}$Ba$_{0.125}$CuO$_4$, our work represents a comprehensive study of the 1/8 anomaly in La-214 cuprates, where superconducting $T_c$ varies by nearly an order of magnitude.

These results are interpreted within a framework of electronic phase separation between stripe regions, which exhibit an inhomogeneous 2$D$ PDW at low temperatures, and homogeneous, hole-disordered regions that display uniform $d$-wave superconductivity. The phase separation is tuned by isovalent Nd$^{3+}$ doping on the La$^{3+}$ site, which induces an LTO-LTT structural phase transition for $y > 0.1$, resulting in samples with predominantly PDW character (Nd-0.4) compared to those with a majority uniform $d$-SC component (LSCO and Nd-0.1), which have much higher 3$D$ superconducting transition temperatures.

Notably, our elastic neutron scattering measurements reveal two surprising results. First, in the Nd-doped samples (Nd-0.1, Nd-0.2, and Nd-0.4), the weak high-temperature background signal at the incommensurate spin stripe wavevector is temperature-dependent, while no such spin stripe signal above $T_N$ is observed for LSCO at $x=1/8$. This indicates that the background in the Nd-containing samples originates from short-range spin stripe correlations that persist to temperatures at least twice $T_N$, and likely throughout the pseudogap phase, as that is the next-highest temperature scale in the problem. This weak, high-temperature signal scales in response to the Nd-concentration in the crystals, and, as mentioned, is not present in LSCO. Second, the LSCO $x=1/8$ spin stripe order parameter evolves in two distinct steps: an initial onset at $T_N$ = 40 K, almost coincident with $T_c^{(onset)}$, with saturation near $T_c^{(mid)}$ = 28 K, followed by a sharp linear increase below $T_\text{PDW}$ = 17 K, where static magnetism is observed via $\mu$SR. We associate this sharp change with the formation of PDW order, which requires spin stripe correlation lengths exceeding roughly 3 $\times$ $\lambda_\text{PDW}$.

\section{Acknowledgment}

We gratefully acknowledge discussions with Stephen Hayden, Steven Kivelson, Young Lee, Subir Sachdev and John M. Tranquada. This work was supported by the Natural Sciences and Engineering Research Council of Canada. A portion of this research used resources at the High Flux Isotope Reactor, a DOE Office of Science User Facility operated by the Oak Ridge National Laboratory.

\bibliography{LSCO}

\begin{thebibliography}{52}%
\makeatletter
\providecommand \@ifxundefined [1]{%
 \@ifx{#1\undefined}
}%
\providecommand \@ifnum [1]{%
 \ifnum #1\expandafter \@firstoftwo
 \else \expandafter \@secondoftwo
 \fi
}%
\providecommand \@ifx [1]{%
 \ifx #1\expandafter \@firstoftwo
 \else \expandafter \@secondoftwo
 \fi
}%
\providecommand \natexlab [1]{#1}%
\providecommand \enquote  [1]{``#1''}%
\providecommand \bibnamefont  [1]{#1}%
\providecommand \bibfnamefont [1]{#1}%
\providecommand \citenamefont [1]{#1}%
\providecommand \href@noop [0]{\@secondoftwo}%
\providecommand \href [0]{\begingroup \@sanitize@url \@href}%
\providecommand \@href[1]{\@@startlink{#1}\@@href}%
\providecommand \@@href[1]{\endgroup#1\@@endlink}%
\providecommand \@sanitize@url [0]{\catcode `\\12\catcode `\$12\catcode `\&12\catcode `\#12\catcode `\^12\catcode `\_12\catcode `\%12\relax}%
\providecommand \@@startlink[1]{}%
\providecommand \@@endlink[0]{}%
\providecommand \url  [0]{\begingroup\@sanitize@url \@url }%
\providecommand \@url [1]{\endgroup\@href {#1}{\urlprefix }}%
\providecommand \urlprefix  [0]{URL }%
\providecommand \Eprint [0]{\href }%
\providecommand \doibase [0]{https://doi.org/}%
\providecommand \selectlanguage [0]{\@gobble}%
\providecommand \bibinfo  [0]{\@secondoftwo}%
\providecommand \bibfield  [0]{\@secondoftwo}%
\providecommand \translation [1]{[#1]}%
\providecommand \BibitemOpen [0]{}%
\providecommand \bibitemStop [0]{}%
\providecommand \bibitemNoStop [0]{.\EOS\space}%
\providecommand \EOS [0]{\spacefactor3000\relax}%
\providecommand \BibitemShut  [1]{\csname bibitem#1\endcsname}%
\let\auto@bib@innerbib\@empty
\bibitem [{\citenamefont {Fradkin}\ \emph {et~al.}(2015)\citenamefont {Fradkin}, \citenamefont {Kivelson},\ and\ \citenamefont {Tranquada}}]{15_fradkin}%
  \BibitemOpen
  \bibfield  {author} {\bibinfo {author} {\bibfnamefont {E.}~\bibnamefont {Fradkin}}, \bibinfo {author} {\bibfnamefont {S.~A.}\ \bibnamefont {Kivelson}},\ and\ \bibinfo {author} {\bibfnamefont {J.~M.}\ \bibnamefont {Tranquada}},\ }\bibfield  {title} {\bibinfo {title} {{Colloquium: Theory of Intertwined Orders in High Temperature Superconductors}},\ }\href {https://doi.org/10.1103/RevModPhys.87.457} {\bibfield  {journal} {\bibinfo  {journal} {Rev. Mod. Phys.}\ }\textbf {\bibinfo {volume} {87}},\ \bibinfo {pages} {457–482} (\bibinfo {year} {2015})}\BibitemShut {NoStop}%
\bibitem [{\citenamefont {Keimer}\ \emph {et~al.}(2015)\citenamefont {Keimer}, \citenamefont {Kivelson}, \citenamefont {Norman}, \citenamefont {Uchida},\ and\ \citenamefont {Zaanen}}]{15_keimer}%
  \BibitemOpen
  \bibfield  {author} {\bibinfo {author} {\bibfnamefont {B.}~\bibnamefont {Keimer}}, \bibinfo {author} {\bibfnamefont {S.~A.}\ \bibnamefont {Kivelson}}, \bibinfo {author} {\bibfnamefont {M.~R.}\ \bibnamefont {Norman}}, \bibinfo {author} {\bibfnamefont {S.}~\bibnamefont {Uchida}},\ and\ \bibinfo {author} {\bibfnamefont {J.}~\bibnamefont {Zaanen}},\ }\bibfield  {title} {\bibinfo {title} {{From quantum matter to high-temperature superconductivity in copper oxides}},\ }\href {https://doi.org/10.1038/nature14165} {\bibfield  {journal} {\bibinfo  {journal} {Nature}\ }\textbf {\bibinfo {volume} {518}},\ \bibinfo {pages} {179–186} (\bibinfo {year} {2015})}\BibitemShut {NoStop}%
\bibitem [{\citenamefont {Proust}\ and\ \citenamefont {Taillefer}(2019)}]{19_proust}%
  \BibitemOpen
  \bibfield  {author} {\bibinfo {author} {\bibfnamefont {C.}~\bibnamefont {Proust}}\ and\ \bibinfo {author} {\bibfnamefont {L.}~\bibnamefont {Taillefer}},\ }\bibfield  {title} {\bibinfo {title} {{The Remarkable Underlying Ground States of Cuprate Superconductors}},\ }\href {https://doi.org/10.1146/annurev-conmatphys-031218-013210} {\bibfield  {journal} {\bibinfo  {journal} {Annual Review of Condensed Matter Physics}\ }\textbf {\bibinfo {volume} {10}},\ \bibinfo {pages} {409–429} (\bibinfo {year} {2019})}\BibitemShut {NoStop}%
\bibitem [{\citenamefont {Agterberg}\ \emph {et~al.}(2020)\citenamefont {Agterberg}, \citenamefont {Davis}, \citenamefont {Edkins}, \citenamefont {Fradkin}, \citenamefont {Van~Harlingen}, \citenamefont {Kivelson}, \citenamefont {Lee}, \citenamefont {Radzihovsky}, \citenamefont {Tranquada},\ and\ \citenamefont {Wang}}]{20_agterberg}%
  \BibitemOpen
  \bibfield  {author} {\bibinfo {author} {\bibfnamefont {D.~F.}\ \bibnamefont {Agterberg}}, \bibinfo {author} {\bibfnamefont {J.~S.}\ \bibnamefont {Davis}}, \bibinfo {author} {\bibfnamefont {S.~D.}\ \bibnamefont {Edkins}}, \bibinfo {author} {\bibfnamefont {E.}~\bibnamefont {Fradkin}}, \bibinfo {author} {\bibfnamefont {D.~J.}\ \bibnamefont {Van~Harlingen}}, \bibinfo {author} {\bibfnamefont {S.~A.}\ \bibnamefont {Kivelson}}, \bibinfo {author} {\bibfnamefont {P.~A.}\ \bibnamefont {Lee}}, \bibinfo {author} {\bibfnamefont {L.}~\bibnamefont {Radzihovsky}}, \bibinfo {author} {\bibfnamefont {J.~M.}\ \bibnamefont {Tranquada}},\ and\ \bibinfo {author} {\bibfnamefont {Y.}~\bibnamefont {Wang}},\ }\bibfield  {title} {\bibinfo {title} {{The Physics of Pair-Density Waves: Cuprate Superconductors and Beyond}},\ }\href {https://doi.org/10.1146/annurev-conmatphys-031119-050711} {\bibfield  {journal} {\bibinfo  {journal} {Annual Review of Condensed Matter Physics}\ }\textbf {\bibinfo {volume} {11}},\ \bibinfo {pages}
  {231–270} (\bibinfo {year} {2020})}\BibitemShut {NoStop}%
\bibitem [{\citenamefont {Berg}\ \emph {et~al.}(2007)\citenamefont {Berg}, \citenamefont {Fradkin}, \citenamefont {Kim}, \citenamefont {Kivelson}, \citenamefont {Oganesyan}, \citenamefont {Tranquada},\ and\ \citenamefont {Zhang}}]{07_berg}%
  \BibitemOpen
  \bibfield  {author} {\bibinfo {author} {\bibfnamefont {E.}~\bibnamefont {Berg}}, \bibinfo {author} {\bibfnamefont {E.}~\bibnamefont {Fradkin}}, \bibinfo {author} {\bibfnamefont {E.-A.}\ \bibnamefont {Kim}}, \bibinfo {author} {\bibfnamefont {S.~A.}\ \bibnamefont {Kivelson}}, \bibinfo {author} {\bibfnamefont {V.}~\bibnamefont {Oganesyan}}, \bibinfo {author} {\bibfnamefont {J.~M.}\ \bibnamefont {Tranquada}},\ and\ \bibinfo {author} {\bibfnamefont {S.~C.}\ \bibnamefont {Zhang}},\ }\bibfield  {title} {\bibinfo {title} {{Dynamical Layer Decoupling in a Stripe-Ordered High-$T_c$ Superconductor}},\ }\href {https://doi.org/10.1103/PhysRevLett.99.127003} {\bibfield  {journal} {\bibinfo  {journal} {Physical Review Letters}\ }\textbf {\bibinfo {volume} {99}},\ \bibinfo {pages} {127003} (\bibinfo {year} {2007})}\BibitemShut {NoStop}%
\bibitem [{\citenamefont {Berg}\ \emph {et~al.}(2009)\citenamefont {Berg}, \citenamefont {Fradkin},\ and\ \citenamefont {Kivelson}}]{09_berg}%
  \BibitemOpen
  \bibfield  {author} {\bibinfo {author} {\bibfnamefont {E.}~\bibnamefont {Berg}}, \bibinfo {author} {\bibfnamefont {E.}~\bibnamefont {Fradkin}},\ and\ \bibinfo {author} {\bibfnamefont {S.~A.}\ \bibnamefont {Kivelson}},\ }\bibfield  {title} {\bibinfo {title} {Theory of the striped superconductor},\ }\href {https://doi.org/10.1103/PhysRevB.79.064515} {\bibfield  {journal} {\bibinfo  {journal} {Physical Review B}\ }\textbf {\bibinfo {volume} {79}},\ \bibinfo {pages} {064515} (\bibinfo {year} {2009})}\BibitemShut {NoStop}%
\bibitem [{\citenamefont {Agterberg}\ and\ \citenamefont {Garaud}(2015)}]{15_agterberg}%
  \BibitemOpen
  \bibfield  {author} {\bibinfo {author} {\bibfnamefont {D.~F.}\ \bibnamefont {Agterberg}}\ and\ \bibinfo {author} {\bibfnamefont {J.}~\bibnamefont {Garaud}},\ }\bibfield  {title} {\bibinfo {title} {{Checkerboard order in vortex cores from pair-density-wave superconductivity}},\ }\href {https://doi.org/10.1103/PhysRevB.91.104512} {\bibfield  {journal} {\bibinfo  {journal} {Physical Review B}\ }\textbf {\bibinfo {volume} {91}},\ \bibinfo {pages} {104512} (\bibinfo {year} {2015})}\BibitemShut {NoStop}%
\bibitem [{\citenamefont {Li}\ \emph {et~al.}(2007)\citenamefont {Li}, \citenamefont {Hücker}, \citenamefont {Gu}, \citenamefont {Tsvelik},\ and\ \citenamefont {Tranquada}}]{07_li}%
  \BibitemOpen
  \bibfield  {author} {\bibinfo {author} {\bibfnamefont {Q.}~\bibnamefont {Li}}, \bibinfo {author} {\bibfnamefont {M.}~\bibnamefont {Hücker}}, \bibinfo {author} {\bibfnamefont {G.~D.}\ \bibnamefont {Gu}}, \bibinfo {author} {\bibfnamefont {A.~M.}\ \bibnamefont {Tsvelik}},\ and\ \bibinfo {author} {\bibfnamefont {J.~M.}\ \bibnamefont {Tranquada}},\ }\bibfield  {title} {\bibinfo {title} {{Two-Dimensional Superconducting Fluctuations in Stripe-Ordered \ch{La_{1.875}Ba_{0.125}CuO4}}},\ }\href {https://doi.org/10.1103/PhysRevLett.99.067001} {\bibfield  {journal} {\bibinfo  {journal} {Physical Review Letters}\ }\textbf {\bibinfo {volume} {99}},\ \bibinfo {pages} {067001} (\bibinfo {year} {2007})}\BibitemShut {NoStop}%
\bibitem [{\citenamefont {Tranquada}\ \emph {et~al.}(2021)\citenamefont {Tranquada}, \citenamefont {Dean},\ and\ \citenamefont {Li}}]{21_tranquada}%
  \BibitemOpen
  \bibfield  {author} {\bibinfo {author} {\bibfnamefont {J.~M.}\ \bibnamefont {Tranquada}}, \bibinfo {author} {\bibfnamefont {M.~P.~M.}\ \bibnamefont {Dean}},\ and\ \bibinfo {author} {\bibfnamefont {Q.}~\bibnamefont {Li}},\ }\bibfield  {title} {\bibinfo {title} {{Superconductivity from Charge Order in Cuprates}},\ }\href {https://doi.org/10.7566/JPSJ.90.111002} {\bibfield  {journal} {\bibinfo  {journal} {J. Phys. Soc. Jpn.}\ }\textbf {\bibinfo {volume} {90}},\ \bibinfo {pages} {111002} (\bibinfo {year} {2021})}\BibitemShut {NoStop}%
\bibitem [{\citenamefont {Huang}\ \emph {et~al.}(2021)\citenamefont {Huang}, \citenamefont {Lee}, \citenamefont {Ikeda}, \citenamefont {Taniguchi}, \citenamefont {Takahama}, \citenamefont {Kao}, \citenamefont {Fujita},\ and\ \citenamefont {Lee}}]{21_huang}%
  \BibitemOpen
  \bibfield  {author} {\bibinfo {author} {\bibfnamefont {H.}~\bibnamefont {Huang}}, \bibinfo {author} {\bibfnamefont {S.-J.}\ \bibnamefont {Lee}}, \bibinfo {author} {\bibfnamefont {Y.}~\bibnamefont {Ikeda}}, \bibinfo {author} {\bibfnamefont {T.}~\bibnamefont {Taniguchi}}, \bibinfo {author} {\bibfnamefont {M.}~\bibnamefont {Takahama}}, \bibinfo {author} {\bibfnamefont {C.-C.}\ \bibnamefont {Kao}}, \bibinfo {author} {\bibfnamefont {M.}~\bibnamefont {Fujita}},\ and\ \bibinfo {author} {\bibfnamefont {J.-S.}\ \bibnamefont {Lee}},\ }\bibfield  {title} {\bibinfo {title} {{Two-Dimensional Superconducting Fluctuations Associated with Charge-Density-Wave Stripes in \ch{La_{1.87}Sr_{0.13}Cu_{0.99}Fe_{0.01}O4}}},\ }\href {https://doi.org/10.1103/PhysRevLett.126.167001} {\bibfield  {journal} {\bibinfo  {journal} {Physical Review Letters}\ }\textbf {\bibinfo {volume} {126}},\ \bibinfo {pages} {167001} (\bibinfo {year} {2021})}\BibitemShut {NoStop}%
\bibitem [{\citenamefont {Wang}\ \emph {et~al.}(2021)\citenamefont {Wang}, \citenamefont {Choubey}, \citenamefont {Chong}, \citenamefont {Chen}, \citenamefont {Ren}, \citenamefont {Eisaki}, \citenamefont {Uchida}, \citenamefont {Hirschfeld},\ and\ \citenamefont {Davis}}]{21_wang}%
  \BibitemOpen
  \bibfield  {author} {\bibinfo {author} {\bibfnamefont {S.}~\bibnamefont {Wang}}, \bibinfo {author} {\bibfnamefont {P.}~\bibnamefont {Choubey}}, \bibinfo {author} {\bibfnamefont {Y.~X.}\ \bibnamefont {Chong}}, \bibinfo {author} {\bibfnamefont {W.}~\bibnamefont {Chen}}, \bibinfo {author} {\bibfnamefont {W.}~\bibnamefont {Ren}}, \bibinfo {author} {\bibfnamefont {H.}~\bibnamefont {Eisaki}}, \bibinfo {author} {\bibfnamefont {S.}~\bibnamefont {Uchida}}, \bibinfo {author} {\bibfnamefont {P.~J.}\ \bibnamefont {Hirschfeld}},\ and\ \bibinfo {author} {\bibfnamefont {J.~C.~S.}\ \bibnamefont {Davis}},\ }\bibfield  {title} {\bibinfo {title} {{Scattering interference signature of a pair density wave state in the cuprate pseudogap phase}},\ }\href {https://doi.org/10.1038/s41467-021-26028-x} {\bibfield  {journal} {\bibinfo  {journal} {Nature Communications}\ }\textbf {\bibinfo {volume} {12}},\ \bibinfo {pages} {6087} (\bibinfo {year} {2021})}\BibitemShut {NoStop}%
\bibitem [{\citenamefont {Choi}\ \emph {et~al.}(2024)\citenamefont {Choi}, \citenamefont {Li}, \citenamefont {Nag}, \citenamefont {Pelliciari}, \citenamefont {Robarts}, \citenamefont {Tam}, \citenamefont {Walters}, \citenamefont {Agrestini}, \citenamefont {García-Fernández}, \citenamefont {Song}, \citenamefont {Eisaki}, \citenamefont {Johnston}, \citenamefont {Comin}, \citenamefont {Ding},\ and\ \citenamefont {Zhou}}]{24_choi}%
  \BibitemOpen
  \bibfield  {author} {\bibinfo {author} {\bibfnamefont {J.}~\bibnamefont {Choi}}, \bibinfo {author} {\bibfnamefont {J.}~\bibnamefont {Li}}, \bibinfo {author} {\bibfnamefont {A.}~\bibnamefont {Nag}}, \bibinfo {author} {\bibfnamefont {J.}~\bibnamefont {Pelliciari}}, \bibinfo {author} {\bibfnamefont {H.}~\bibnamefont {Robarts}}, \bibinfo {author} {\bibfnamefont {C.~C.}\ \bibnamefont {Tam}}, \bibinfo {author} {\bibfnamefont {A.}~\bibnamefont {Walters}}, \bibinfo {author} {\bibfnamefont {S.}~\bibnamefont {Agrestini}}, \bibinfo {author} {\bibfnamefont {M.}~\bibnamefont {García-Fernández}}, \bibinfo {author} {\bibfnamefont {D.}~\bibnamefont {Song}}, \bibinfo {author} {\bibfnamefont {H.}~\bibnamefont {Eisaki}}, \bibinfo {author} {\bibfnamefont {S.}~\bibnamefont {Johnston}}, \bibinfo {author} {\bibfnamefont {R.}~\bibnamefont {Comin}}, \bibinfo {author} {\bibfnamefont {H.}~\bibnamefont {Ding}},\ and\ \bibinfo {author} {\bibfnamefont {K.-J.}\ \bibnamefont {Zhou}},\ }\bibfield  {title} {\bibinfo {title} {{Universal
  stripe symmetry of short-range charge density waves in cuprate superconductors}},\ }\href {https://doi.org/10.1002/adma.202307515} {\bibfield  {journal} {\bibinfo  {journal} {Adv. Mater.}\ }\textbf {\bibinfo {volume} {36}},\ \bibinfo {pages} {2307515} (\bibinfo {year} {2024})}\BibitemShut {NoStop}%
\bibitem [{\citenamefont {Himeda}\ \emph {et~al.}(2002)\citenamefont {Himeda}, \citenamefont {Kato},\ and\ \citenamefont {Ogata}}]{02_himeda}%
  \BibitemOpen
  \bibfield  {author} {\bibinfo {author} {\bibfnamefont {A.}~\bibnamefont {Himeda}}, \bibinfo {author} {\bibfnamefont {T.}~\bibnamefont {Kato}},\ and\ \bibinfo {author} {\bibfnamefont {M.}~\bibnamefont {Ogata}},\ }\bibfield  {title} {\bibinfo {title} {{Stripe States with Spatially Oscillating $d$-Wave Superconductivity in the Two-Dimensional $t-t'-J$ Model}},\ }\href {https://doi.org/10.1103/PhysRevLett.88.117001} {\bibfield  {journal} {\bibinfo  {journal} {Physical Review Letters}\ }\textbf {\bibinfo {volume} {88}},\ \bibinfo {pages} {117001} (\bibinfo {year} {2002})}\BibitemShut {NoStop}%
\bibitem [{\citenamefont {Corboz}(2014)}]{14_corboz}%
  \BibitemOpen
  \bibfield  {author} {\bibinfo {author} {\bibfnamefont {P.}~\bibnamefont {Corboz}},\ }\bibfield  {title} {\bibinfo {title} {{Competing States in the $t-J$ Model: Uniform $d$-Wave State versus Stripe State}},\ }\bibfield  {journal} {\bibinfo  {journal} {Physical Review Letters}\ }\textbf {\bibinfo {volume} {113}},\ \href {https://doi.org/10.1103/PhysRevLett.113.046402} {10.1103/PhysRevLett.113.046402} (\bibinfo {year} {2014})\BibitemShut {NoStop}%
\bibitem [{\citenamefont {Choubey}\ \emph {et~al.}(2020)\citenamefont {Choubey}, \citenamefont {Joo}, \citenamefont {Fujita}, \citenamefont {Du}, \citenamefont {Edkins}, \citenamefont {Hamidian}, \citenamefont {Eisaki}, \citenamefont {Uchida}, \citenamefont {Mackenzie}, \citenamefont {Lee}, \citenamefont {Davis},\ and\ \citenamefont {Hirschfeld}}]{20_choubey}%
  \BibitemOpen
  \bibfield  {author} {\bibinfo {author} {\bibfnamefont {P.}~\bibnamefont {Choubey}}, \bibinfo {author} {\bibfnamefont {S.~H.}\ \bibnamefont {Joo}}, \bibinfo {author} {\bibfnamefont {K.}~\bibnamefont {Fujita}}, \bibinfo {author} {\bibfnamefont {Z.}~\bibnamefont {Du}}, \bibinfo {author} {\bibfnamefont {S.~D.}\ \bibnamefont {Edkins}}, \bibinfo {author} {\bibfnamefont {M.~H.}\ \bibnamefont {Hamidian}}, \bibinfo {author} {\bibfnamefont {H.}~\bibnamefont {Eisaki}}, \bibinfo {author} {\bibfnamefont {S.}~\bibnamefont {Uchida}}, \bibinfo {author} {\bibfnamefont {A.~P.}\ \bibnamefont {Mackenzie}}, \bibinfo {author} {\bibfnamefont {J.}~\bibnamefont {Lee}}, \bibinfo {author} {\bibfnamefont {J.~C.~S.}\ \bibnamefont {Davis}},\ and\ \bibinfo {author} {\bibfnamefont {P.~J.}\ \bibnamefont {Hirschfeld}},\ }\bibfield  {title} {\bibinfo {title} {{Atomic-scale electronic structure of the cuprate pair density wave state coexisting with superconductivity}},\ }\href {https://doi.org/10.1073/pnas.2002429117} {\bibfield  {journal}
  {\bibinfo  {journal} {Proceedings of the National Academy of Sciences}\ }\textbf {\bibinfo {volume} {117}},\ \bibinfo {pages} {14805–14811} (\bibinfo {year} {2020})}\BibitemShut {NoStop}%
\bibitem [{\citenamefont {Setty}\ \emph {et~al.}(2023)\citenamefont {Setty}, \citenamefont {Fanfarillo},\ and\ \citenamefont {Hirschfeld}}]{23_setty}%
  \BibitemOpen
  \bibfield  {author} {\bibinfo {author} {\bibfnamefont {C.}~\bibnamefont {Setty}}, \bibinfo {author} {\bibfnamefont {L.}~\bibnamefont {Fanfarillo}},\ and\ \bibinfo {author} {\bibfnamefont {P.~J.}\ \bibnamefont {Hirschfeld}},\ }\bibfield  {title} {\bibinfo {title} {{Mechanism for fluctuating pair density wave}},\ }\href {https://doi.org/10.1038/s41467-023-38956-x} {\bibfield  {journal} {\bibinfo  {journal} {Nature Communications}\ }\textbf {\bibinfo {volume} {14}},\ \bibinfo {pages} {3181} (\bibinfo {year} {2023})}\BibitemShut {NoStop}%
\bibitem [{\citenamefont {Yamada}\ \emph {et~al.}(1998)\citenamefont {Yamada}, \citenamefont {Lee}, \citenamefont {Kurahashi}, \citenamefont {Wada}, \citenamefont {Wakimoto}, \citenamefont {Ueki}, \citenamefont {Kimura}, \citenamefont {Endoh}, \citenamefont {Hosoya}, \citenamefont {Shirane}, \citenamefont {Birgeneau}, \citenamefont {Greven}, \citenamefont {Kastner},\ and\ \citenamefont {Kim}}]{98_yamada}%
  \BibitemOpen
  \bibfield  {author} {\bibinfo {author} {\bibfnamefont {K.}~\bibnamefont {Yamada}}, \bibinfo {author} {\bibfnamefont {C.~H.}\ \bibnamefont {Lee}}, \bibinfo {author} {\bibfnamefont {K.}~\bibnamefont {Kurahashi}}, \bibinfo {author} {\bibfnamefont {J.}~\bibnamefont {Wada}}, \bibinfo {author} {\bibfnamefont {S.}~\bibnamefont {Wakimoto}}, \bibinfo {author} {\bibfnamefont {S.}~\bibnamefont {Ueki}}, \bibinfo {author} {\bibfnamefont {H.}~\bibnamefont {Kimura}}, \bibinfo {author} {\bibfnamefont {Y.}~\bibnamefont {Endoh}}, \bibinfo {author} {\bibfnamefont {S.}~\bibnamefont {Hosoya}}, \bibinfo {author} {\bibfnamefont {G.}~\bibnamefont {Shirane}}, \bibinfo {author} {\bibfnamefont {R.~J.}\ \bibnamefont {Birgeneau}}, \bibinfo {author} {\bibfnamefont {M.}~\bibnamefont {Greven}}, \bibinfo {author} {\bibfnamefont {M.~A.}\ \bibnamefont {Kastner}},\ and\ \bibinfo {author} {\bibfnamefont {Y.~J.}\ \bibnamefont {Kim}},\ }\bibfield  {title} {\bibinfo {title} {{Doping Dependence of the Spatially Modulated Dynamical Spin
  Correlations and the Superconducting-transition Temperature in \ch{La_{2-$x$}Sr_{$x$}CuO4}}},\ }\href {https://doi.org/10.1103/PhysRevB.57.6165} {\bibfield  {journal} {\bibinfo  {journal} {Phys. Rev. B}\ }\textbf {\bibinfo {volume} {57}},\ \bibinfo {pages} {6165–6172} (\bibinfo {year} {1998})}\BibitemShut {NoStop}%
\bibitem [{\citenamefont {Dragomir}\ \emph {et~al.}(2020)\citenamefont {Dragomir}, \citenamefont {Ma}, \citenamefont {Clancy}, \citenamefont {Ataei}, \citenamefont {Dube}, \citenamefont {Sharma}, \citenamefont {Huq}, \citenamefont {Dabkowska}, \citenamefont {Taillefer},\ and\ \citenamefont {Gaulin}}]{20_dragomir}%
  \BibitemOpen
  \bibfield  {author} {\bibinfo {author} {\bibfnamefont {M.}~\bibnamefont {Dragomir}}, \bibinfo {author} {\bibfnamefont {Q.}~\bibnamefont {Ma}}, \bibinfo {author} {\bibfnamefont {J.~P.}\ \bibnamefont {Clancy}}, \bibinfo {author} {\bibfnamefont {A.}~\bibnamefont {Ataei}}, \bibinfo {author} {\bibfnamefont {P.~A.}\ \bibnamefont {Dube}}, \bibinfo {author} {\bibfnamefont {S.}~\bibnamefont {Sharma}}, \bibinfo {author} {\bibfnamefont {A.}~\bibnamefont {Huq}}, \bibinfo {author} {\bibfnamefont {H.~A.}\ \bibnamefont {Dabkowska}}, \bibinfo {author} {\bibfnamefont {L.}~\bibnamefont {Taillefer}},\ and\ \bibinfo {author} {\bibfnamefont {B.~D.}\ \bibnamefont {Gaulin}},\ }\bibfield  {title} {\bibinfo {title} {{Materials preparation, single-crystal growth, and the phase diagram of the cuprate high-temperature superconductor \ch{La_{1.6-$x$}Nd_{0.4}Sr_{$x$}CuO4}}},\ }\href {https://doi.org/10.1103/PhysRevMaterials.4.114801} {\bibfield  {journal} {\bibinfo  {journal} {Phys. Rev. Mater.}\ }\textbf {\bibinfo {volume} {4}},\
  \bibinfo {pages} {114801} (\bibinfo {year} {2020})}\BibitemShut {NoStop}%
\bibitem [{\citenamefont {Ma}\ \emph {et~al.}(2022)\citenamefont {Ma}, \citenamefont {Smith}, \citenamefont {Cronkwright}, \citenamefont {Dragomir}, \citenamefont {Mitchell}, \citenamefont {Winn}, \citenamefont {Williams},\ and\ \citenamefont {Gaulin}}]{22_ma2}%
  \BibitemOpen
  \bibfield  {author} {\bibinfo {author} {\bibfnamefont {Q.}~\bibnamefont {Ma}}, \bibinfo {author} {\bibfnamefont {E.~M.}\ \bibnamefont {Smith}}, \bibinfo {author} {\bibfnamefont {Z.~W.}\ \bibnamefont {Cronkwright}}, \bibinfo {author} {\bibfnamefont {M.}~\bibnamefont {Dragomir}}, \bibinfo {author} {\bibfnamefont {G.}~\bibnamefont {Mitchell}}, \bibinfo {author} {\bibfnamefont {B.~W.}\ \bibnamefont {Winn}}, \bibinfo {author} {\bibfnamefont {T.~J.}\ \bibnamefont {Williams}},\ and\ \bibinfo {author} {\bibfnamefont {B.~D.}\ \bibnamefont {Gaulin}},\ }\bibfield  {title} {\bibinfo {title} {{Magnetic field tuning of parallel spin stripe order and fluctuations near the pseudogap quantum critical point in \ch{La_{1.36}Nd_{0.4}Sr_{0.24}CuO4}}},\ }\href {https://doi.org/10.1103/PhysRevB.106.214427} {\bibfield  {journal} {\bibinfo  {journal} {Phys. Rev. B}\ }\textbf {\bibinfo {volume} {106}},\ \bibinfo {pages} {214427} (\bibinfo {year} {2022})}\BibitemShut {NoStop}%
\bibitem [{\citenamefont {Crawford}\ \emph {et~al.}(1991)\citenamefont {Crawford}, \citenamefont {Harlow}, \citenamefont {McCarron}, \citenamefont {Farneth}, \citenamefont {Axe}, \citenamefont {Chou},\ and\ \citenamefont {Huang}}]{91_crawford}%
  \BibitemOpen
  \bibfield  {author} {\bibinfo {author} {\bibfnamefont {M.~K.}\ \bibnamefont {Crawford}}, \bibinfo {author} {\bibfnamefont {R.~L.}\ \bibnamefont {Harlow}}, \bibinfo {author} {\bibfnamefont {E.~M.}\ \bibnamefont {McCarron}}, \bibinfo {author} {\bibfnamefont {W.~E.}\ \bibnamefont {Farneth}}, \bibinfo {author} {\bibfnamefont {J.~D.}\ \bibnamefont {Axe}}, \bibinfo {author} {\bibfnamefont {H.}~\bibnamefont {Chou}},\ and\ \bibinfo {author} {\bibfnamefont {Q.}~\bibnamefont {Huang}},\ }\bibfield  {title} {\bibinfo {title} {{Lattice instabilities and the effect of copper-oxygen-sheet distortions on superconductivity in doped \ch{La2CuO4}}},\ }\href {https://doi.org/10.1103/PhysRevB.44.7749} {\bibfield  {journal} {\bibinfo  {journal} {Phys. Rev. B}\ }\textbf {\bibinfo {volume} {44}},\ \bibinfo {pages} {7749} (\bibinfo {year} {1991})}\BibitemShut {NoStop}%
\bibitem [{\citenamefont {Tajima}\ \emph {et~al.}(2001)\citenamefont {Tajima}, \citenamefont {Noda}, \citenamefont {Eisaki},\ and\ \citenamefont {Uchida}}]{01_tajima}%
  \BibitemOpen
  \bibfield  {author} {\bibinfo {author} {\bibfnamefont {S.}~\bibnamefont {Tajima}}, \bibinfo {author} {\bibfnamefont {T.}~\bibnamefont {Noda}}, \bibinfo {author} {\bibfnamefont {H.}~\bibnamefont {Eisaki}},\ and\ \bibinfo {author} {\bibfnamefont {S.}~\bibnamefont {Uchida}},\ }\bibfield  {title} {\bibinfo {title} {{$c$-Axis optical response in the static stripe ordered phase of the cuprates}},\ }\href {https://doi.org/10.1103/PhysRevLett.86.500} {\bibfield  {journal} {\bibinfo  {journal} {Phys. Rev. Lett.}\ }\textbf {\bibinfo {volume} {86}},\ \bibinfo {pages} {500} (\bibinfo {year} {2001})}\BibitemShut {NoStop}%
\bibitem [{\citenamefont {Suzuki}(2012)}]{12_suzuki}%
  \BibitemOpen
  \bibfield  {author} {\bibinfo {author} {\bibfnamefont {K.~M.}\ \bibnamefont {Suzuki}},\ }\bibfield  {title} {\bibinfo {title} {{Distinct Fe-induced magnetic states in the underdoped and overdoped regimes of \ch{La_{2-x}Sr_{x}Cu_{1-y}Fe_{y}O4} revealed by muon spin relaxation}},\ }\bibfield  {journal} {\bibinfo  {journal} {Physical Review B}\ }\textbf {\bibinfo {volume} {86}},\ \href {https://doi.org/10.1103/PhysRevB.86.014522} {10.1103/PhysRevB.86.014522} (\bibinfo {year} {2012})\BibitemShut {NoStop}%
\bibitem [{\citenamefont {Lee}\ \emph {et~al.}(1999)\citenamefont {Lee}, \citenamefont {Birgeneau}, \citenamefont {Kastner}, \citenamefont {Endoh}, \citenamefont {Wakimoto}, \citenamefont {Yamada}, \citenamefont {Erwin}, \citenamefont {Lee},\ and\ \citenamefont {Shirane}}]{99_Lee}%
  \BibitemOpen
  \bibfield  {author} {\bibinfo {author} {\bibfnamefont {Y.~S.}\ \bibnamefont {Lee}}, \bibinfo {author} {\bibfnamefont {R.~J.}\ \bibnamefont {Birgeneau}}, \bibinfo {author} {\bibfnamefont {M.~A.}\ \bibnamefont {Kastner}}, \bibinfo {author} {\bibfnamefont {Y.}~\bibnamefont {Endoh}}, \bibinfo {author} {\bibfnamefont {S.}~\bibnamefont {Wakimoto}}, \bibinfo {author} {\bibfnamefont {K.}~\bibnamefont {Yamada}}, \bibinfo {author} {\bibfnamefont {R.~W.}\ \bibnamefont {Erwin}}, \bibinfo {author} {\bibfnamefont {S.-H.}\ \bibnamefont {Lee}},\ and\ \bibinfo {author} {\bibfnamefont {G.}~\bibnamefont {Shirane}},\ }\bibfield  {title} {\bibinfo {title} {{Neutron-scattering study of spin-density wave order in the superconducting state of excess-oxygen-doped \ch{La2CuO_{4+$y$}}}},\ }\href {https://doi.org/10.1103/PhysRevB.60.3643} {\bibfield  {journal} {\bibinfo  {journal} {Phys. Rev. B}\ }\textbf {\bibinfo {volume} {60}},\ \bibinfo {pages} {3643–3654} (\bibinfo {year} {1999})}\BibitemShut {NoStop}%
\bibitem [{\citenamefont {Kimura}\ \emph {et~al.}(2000)\citenamefont {Kimura}, \citenamefont {Matsushita}, \citenamefont {Hirota}, \citenamefont {Endoh}, \citenamefont {Yamada}, \citenamefont {Shirane}, \citenamefont {Lee}, \citenamefont {Kastner},\ and\ \citenamefont {Birgeneau}}]{00_kimura}%
  \BibitemOpen
  \bibfield  {author} {\bibinfo {author} {\bibfnamefont {H.}~\bibnamefont {Kimura}}, \bibinfo {author} {\bibfnamefont {H.}~\bibnamefont {Matsushita}}, \bibinfo {author} {\bibfnamefont {K.}~\bibnamefont {Hirota}}, \bibinfo {author} {\bibfnamefont {Y.}~\bibnamefont {Endoh}}, \bibinfo {author} {\bibfnamefont {K.}~\bibnamefont {Yamada}}, \bibinfo {author} {\bibfnamefont {G.}~\bibnamefont {Shirane}}, \bibinfo {author} {\bibfnamefont {Y.~S.}\ \bibnamefont {Lee}}, \bibinfo {author} {\bibfnamefont {M.~A.}\ \bibnamefont {Kastner}},\ and\ \bibinfo {author} {\bibfnamefont {R.~J.}\ \bibnamefont {Birgeneau}},\ }\bibfield  {title} {\bibinfo {title} {{Incommensurate geometry of the elastic magnetic peaks in superconducting \ch{La_{1.88}Sr_{0.12}CuO4}}},\ }\href {https://doi.org/10.1103/PhysRevB.61.14366} {\bibfield  {journal} {\bibinfo  {journal} {Phys. Rev. B}\ }\textbf {\bibinfo {volume} {61}},\ \bibinfo {pages} {14366} (\bibinfo {year} {2000})}\BibitemShut {NoStop}%
\bibitem [{\citenamefont {Katano}\ \emph {et~al.}(2000)\citenamefont {Katano}, \citenamefont {Sato}, \citenamefont {Yamada}, \citenamefont {Suzuki},\ and\ \citenamefont {Fukase}}]{00_Katano}%
  \BibitemOpen
  \bibfield  {author} {\bibinfo {author} {\bibfnamefont {S.}~\bibnamefont {Katano}}, \bibinfo {author} {\bibfnamefont {M.}~\bibnamefont {Sato}}, \bibinfo {author} {\bibfnamefont {K.}~\bibnamefont {Yamada}}, \bibinfo {author} {\bibfnamefont {T.}~\bibnamefont {Suzuki}},\ and\ \bibinfo {author} {\bibfnamefont {T.}~\bibnamefont {Fukase}},\ }\bibfield  {title} {\bibinfo {title} {{Enhancement of static antiferromagnetic correlations by magnetic field in a superconductor \ch{La_{2-$x$}Sr_{$x$}CuO4} with $x$ = 0.12}},\ }\href {https://doi.org/10.1103/PhysRevB.62.R14677} {\bibfield  {journal} {\bibinfo  {journal} {Phys. Rev. B}\ }\textbf {\bibinfo {volume} {62}},\ \bibinfo {pages} {R14677} (\bibinfo {year} {2000})}\BibitemShut {NoStop}%
\bibitem [{\citenamefont {Fujita}\ \emph {et~al.}(2002)\citenamefont {Fujita}, \citenamefont {Goka}, \citenamefont {Yamada},\ and\ \citenamefont {Matsuda}}]{02_Fujita}%
  \BibitemOpen
  \bibfield  {author} {\bibinfo {author} {\bibfnamefont {M.}~\bibnamefont {Fujita}}, \bibinfo {author} {\bibfnamefont {H.}~\bibnamefont {Goka}}, \bibinfo {author} {\bibfnamefont {K.}~\bibnamefont {Yamada}},\ and\ \bibinfo {author} {\bibfnamefont {M.}~\bibnamefont {Matsuda}},\ }\bibfield  {title} {\bibinfo {title} {{Competition between charge- and spin-density-wave order and superconductivity in \ch{La_{1.875}Ba_{0.125-$x$}Sr_{$x$}CuO4}}},\ }\href {https://doi.org/10.1103/PhysRevLett.88.167008} {\bibfield  {journal} {\bibinfo  {journal} {Phys. Rev. Lett.}\ }\textbf {\bibinfo {volume} {88}},\ \bibinfo {pages} {167008} (\bibinfo {year} {2002})}\BibitemShut {NoStop}%
\bibitem [{\citenamefont {He}\ \emph {et~al.}(2024)\citenamefont {He}, \citenamefont {Wen}, \citenamefont {Jiang}, \citenamefont {Xu}, \citenamefont {Tian}, \citenamefont {Taniguchi}, \citenamefont {Ikeda}, \citenamefont {Fujita},\ and\ \citenamefont {Lee}}]{24_he}%
  \BibitemOpen
  \bibfield  {author} {\bibinfo {author} {\bibfnamefont {W.}~\bibnamefont {He}}, \bibinfo {author} {\bibfnamefont {J.}~\bibnamefont {Wen}}, \bibinfo {author} {\bibfnamefont {H.-C.}\ \bibnamefont {Jiang}}, \bibinfo {author} {\bibfnamefont {G.}~\bibnamefont {Xu}}, \bibinfo {author} {\bibfnamefont {W.}~\bibnamefont {Tian}}, \bibinfo {author} {\bibfnamefont {T.}~\bibnamefont {Taniguchi}}, \bibinfo {author} {\bibfnamefont {Y.}~\bibnamefont {Ikeda}}, \bibinfo {author} {\bibfnamefont {M.}~\bibnamefont {Fujita}},\ and\ \bibinfo {author} {\bibfnamefont {Y.~S.}\ \bibnamefont {Lee}},\ }\bibfield  {title} {\bibinfo {title} {{Tilted stripes origin in \ch{La_{1.88}Sr_{0.12}CuO4} revealed by anisotropic next-nearest neighbor hopping}},\ }\href {https://doi.org/10.1038/s42005-024-01753-z} {\bibfield  {journal} {\bibinfo  {journal} {Commun. Phys.}\ }\textbf {\bibinfo {volume} {7}},\ \bibinfo {pages} {1–9} (\bibinfo {year} {2024})}\BibitemShut {NoStop}%
\bibitem [{\citenamefont {Wang}\ \emph {et~al.}(2022)\citenamefont {Wang}, \citenamefont {von Arx}, \citenamefont {Mazzone}, \citenamefont {Mustafi}, \citenamefont {Horio}, \citenamefont {Küspert}, \citenamefont {Choi}, \citenamefont {Bucher}, \citenamefont {Wo}, \citenamefont {Zhao}, \citenamefont {Zhang}, \citenamefont {Asmara}, \citenamefont {Sassa}, \citenamefont {Månsson}, \citenamefont {Christensen}, \citenamefont {Janoschek}, \citenamefont {Kurosawa}, \citenamefont {Momono}, \citenamefont {Oda}, \citenamefont {Fischer}, \citenamefont {Schmitt},\ and\ \citenamefont {Chang}}]{22_Wang}%
  \BibitemOpen
  \bibfield  {author} {\bibinfo {author} {\bibfnamefont {Q.}~\bibnamefont {Wang}}, \bibinfo {author} {\bibfnamefont {K.}~\bibnamefont {von Arx}}, \bibinfo {author} {\bibfnamefont {D.~G.}\ \bibnamefont {Mazzone}}, \bibinfo {author} {\bibfnamefont {S.}~\bibnamefont {Mustafi}}, \bibinfo {author} {\bibfnamefont {M.}~\bibnamefont {Horio}}, \bibinfo {author} {\bibfnamefont {J.}~\bibnamefont {Küspert}}, \bibinfo {author} {\bibfnamefont {J.}~\bibnamefont {Choi}}, \bibinfo {author} {\bibfnamefont {D.}~\bibnamefont {Bucher}}, \bibinfo {author} {\bibfnamefont {H.}~\bibnamefont {Wo}}, \bibinfo {author} {\bibfnamefont {J.}~\bibnamefont {Zhao}}, \bibinfo {author} {\bibfnamefont {W.}~\bibnamefont {Zhang}}, \bibinfo {author} {\bibfnamefont {T.~C.}\ \bibnamefont {Asmara}}, \bibinfo {author} {\bibfnamefont {Y.}~\bibnamefont {Sassa}}, \bibinfo {author} {\bibfnamefont {M.}~\bibnamefont {Månsson}}, \bibinfo {author} {\bibfnamefont {N.~B.}\ \bibnamefont {Christensen}}, \bibinfo {author} {\bibfnamefont {M.}~\bibnamefont
  {Janoschek}}, \bibinfo {author} {\bibfnamefont {T.}~\bibnamefont {Kurosawa}}, \bibinfo {author} {\bibfnamefont {N.}~\bibnamefont {Momono}}, \bibinfo {author} {\bibfnamefont {M.}~\bibnamefont {Oda}}, \bibinfo {author} {\bibfnamefont {M.~H.}\ \bibnamefont {Fischer}}, \bibinfo {author} {\bibfnamefont {T.}~\bibnamefont {Schmitt}},\ and\ \bibinfo {author} {\bibfnamefont {J.}~\bibnamefont {Chang}},\ }\bibfield  {title} {\bibinfo {title} {{Uniaxial pressure induced stripe order rotation in \ch{La_{1.88}Sr_{0.12}CuO4}}},\ }\href {https://doi.org/10.1038/s41467-022-29465-4} {\bibfield  {journal} {\bibinfo  {journal} {Nat. Commun.}\ }\textbf {\bibinfo {volume} {13}},\ \bibinfo {pages} {1795} (\bibinfo {year} {2022})}\BibitemShut {NoStop}%
\bibitem [{\citenamefont {Tranquada}\ \emph {et~al.}(1996)\citenamefont {Tranquada}, \citenamefont {Axe}, \citenamefont {Ichikawa}, \citenamefont {Nakamura}, \citenamefont {Uchida},\ and\ \citenamefont {Nachumi}}]{96_tranquada}%
  \BibitemOpen
  \bibfield  {author} {\bibinfo {author} {\bibfnamefont {J.~M.}\ \bibnamefont {Tranquada}}, \bibinfo {author} {\bibfnamefont {J.~D.}\ \bibnamefont {Axe}}, \bibinfo {author} {\bibfnamefont {N.}~\bibnamefont {Ichikawa}}, \bibinfo {author} {\bibfnamefont {Y.}~\bibnamefont {Nakamura}}, \bibinfo {author} {\bibfnamefont {S.}~\bibnamefont {Uchida}},\ and\ \bibinfo {author} {\bibfnamefont {B.}~\bibnamefont {Nachumi}},\ }\bibfield  {title} {\bibinfo {title} {{Neutron scattering study of stripe-phase order of holes and spins in \ch{La_{1.48}Nd_{0.4}Sr_{0.12}CuO4}}},\ }\href {https://doi.org/10.1103/PhysRevB.54.7489} {\bibfield  {journal} {\bibinfo  {journal} {Phys. Rev. B}\ }\textbf {\bibinfo {volume} {54}},\ \bibinfo {pages} {7489–7499} (\bibinfo {year} {1996})}\BibitemShut {NoStop}%
\bibitem [{\citenamefont {Chang}\ \emph {et~al.}(2008)\citenamefont {Chang}, \citenamefont {Niedermayer}, \citenamefont {Gilardi}, \citenamefont {Christensen}, \citenamefont {R\o{}nnow}, \citenamefont {McMorrow}, \citenamefont {Ay}, \citenamefont {Stahn}, \citenamefont {Sobolev}, \citenamefont {Hiess}, \citenamefont {Pailhes}, \citenamefont {Baines}, \citenamefont {Momono}, \citenamefont {Oda}, \citenamefont {Ido},\ and\ \citenamefont {Mesot}}]{08_changj}%
  \BibitemOpen
  \bibfield  {author} {\bibinfo {author} {\bibfnamefont {J.}~\bibnamefont {Chang}}, \bibinfo {author} {\bibfnamefont {C.}~\bibnamefont {Niedermayer}}, \bibinfo {author} {\bibfnamefont {R.}~\bibnamefont {Gilardi}}, \bibinfo {author} {\bibfnamefont {N.~B.}\ \bibnamefont {Christensen}}, \bibinfo {author} {\bibfnamefont {H.~M.}\ \bibnamefont {R\o{}nnow}}, \bibinfo {author} {\bibfnamefont {D.~F.}\ \bibnamefont {McMorrow}}, \bibinfo {author} {\bibfnamefont {M.}~\bibnamefont {Ay}}, \bibinfo {author} {\bibfnamefont {J.}~\bibnamefont {Stahn}}, \bibinfo {author} {\bibfnamefont {O.}~\bibnamefont {Sobolev}}, \bibinfo {author} {\bibfnamefont {A.}~\bibnamefont {Hiess}}, \bibinfo {author} {\bibfnamefont {S.}~\bibnamefont {Pailhes}}, \bibinfo {author} {\bibfnamefont {C.}~\bibnamefont {Baines}}, \bibinfo {author} {\bibfnamefont {N.}~\bibnamefont {Momono}}, \bibinfo {author} {\bibfnamefont {M.}~\bibnamefont {Oda}}, \bibinfo {author} {\bibfnamefont {M.}~\bibnamefont {Ido}},\ and\ \bibinfo {author} {\bibfnamefont
  {J.}~\bibnamefont {Mesot}},\ }\bibfield  {title} {\bibinfo {title} {{Tuning Competing Orders in \ch{La_{2-$x$}Sr_{$x$}CuO4} Cuprate Superconductors by the Application of an External Magnetic Field}},\ }\href {https://doi.org/10.1103/PhysRevB.78.104525} {\bibfield  {journal} {\bibinfo  {journal} {Phys. Rev. B}\ }\textbf {\bibinfo {volume} {78}},\ \bibinfo {pages} {104525} (\bibinfo {year} {2008})}\BibitemShut {NoStop}%
\bibitem [{\citenamefont {Savici}\ \emph {et~al.}(2002)\citenamefont {Savici}, \citenamefont {Fudamoto}, \citenamefont {Gat}, \citenamefont {Ito}, \citenamefont {Larkin}, \citenamefont {Uemura}, \citenamefont {Luke}, \citenamefont {Kojima}, \citenamefont {Lee}, \citenamefont {Kastner}, \citenamefont {Birgeneau},\ and\ \citenamefont {Yamada}}]{02_savici}%
  \BibitemOpen
  \bibfield  {author} {\bibinfo {author} {\bibfnamefont {A.~T.}\ \bibnamefont {Savici}}, \bibinfo {author} {\bibfnamefont {Y.}~\bibnamefont {Fudamoto}}, \bibinfo {author} {\bibfnamefont {I.~M.}\ \bibnamefont {Gat}}, \bibinfo {author} {\bibfnamefont {T.}~\bibnamefont {Ito}}, \bibinfo {author} {\bibfnamefont {M.~I.}\ \bibnamefont {Larkin}}, \bibinfo {author} {\bibfnamefont {Y.~J.}\ \bibnamefont {Uemura}}, \bibinfo {author} {\bibfnamefont {G.~M.}\ \bibnamefont {Luke}}, \bibinfo {author} {\bibfnamefont {K.~M.}\ \bibnamefont {Kojima}}, \bibinfo {author} {\bibfnamefont {Y.~S.}\ \bibnamefont {Lee}}, \bibinfo {author} {\bibfnamefont {M.~A.}\ \bibnamefont {Kastner}}, \bibinfo {author} {\bibfnamefont {R.~J.}\ \bibnamefont {Birgeneau}},\ and\ \bibinfo {author} {\bibfnamefont {K.}~\bibnamefont {Yamada}},\ }\bibfield  {title} {\bibinfo {title} {{Muon spin relaxation studies of incommensurate magnetism and superconductivity in stage-4 \ch{La2CuO_{4.11}} and \ch{La_{1.88}Sr_{0.12}CuO4}}},\ }\href
  {https://doi.org/10.1103/PhysRevB.66.014524} {\bibfield  {journal} {\bibinfo  {journal} {Physical Review B}\ }\textbf {\bibinfo {volume} {66}},\ \bibinfo {pages} {014524} (\bibinfo {year} {2002})}\BibitemShut {NoStop}%
\bibitem [{\citenamefont {Savici}\ \emph {et~al.}(2005)\citenamefont {Savici}, \citenamefont {Fukaya}, \citenamefont {Gat-Malureanu}, \citenamefont {Ito}, \citenamefont {Russo}, \citenamefont {Uemura}, \citenamefont {Wiebe}, \citenamefont {Kyriakou}, \citenamefont {MacDougall}, \citenamefont {Rovers}, \citenamefont {Luke}, \citenamefont {Kojima}, \citenamefont {Goto}, \citenamefont {Uchida}, \citenamefont {Kadono}, \citenamefont {Yamada}, \citenamefont {Tajima}, \citenamefont {Masui}, \citenamefont {Eisaki}, \citenamefont {Kaneko}, \citenamefont {Greven},\ and\ \citenamefont {Gu}}]{05_savici}%
  \BibitemOpen
  \bibfield  {author} {\bibinfo {author} {\bibfnamefont {A.~T.}\ \bibnamefont {Savici}}, \bibinfo {author} {\bibfnamefont {A.}~\bibnamefont {Fukaya}}, \bibinfo {author} {\bibfnamefont {I.~M.}\ \bibnamefont {Gat-Malureanu}}, \bibinfo {author} {\bibfnamefont {T.}~\bibnamefont {Ito}}, \bibinfo {author} {\bibfnamefont {P.~L.}\ \bibnamefont {Russo}}, \bibinfo {author} {\bibfnamefont {Y.~J.}\ \bibnamefont {Uemura}}, \bibinfo {author} {\bibfnamefont {C.~R.}\ \bibnamefont {Wiebe}}, \bibinfo {author} {\bibfnamefont {P.~P.}\ \bibnamefont {Kyriakou}}, \bibinfo {author} {\bibfnamefont {G.~J.}\ \bibnamefont {MacDougall}}, \bibinfo {author} {\bibfnamefont {M.~T.}\ \bibnamefont {Rovers}}, \bibinfo {author} {\bibfnamefont {G.~M.}\ \bibnamefont {Luke}}, \bibinfo {author} {\bibfnamefont {K.~M.}\ \bibnamefont {Kojima}}, \bibinfo {author} {\bibfnamefont {M.}~\bibnamefont {Goto}}, \bibinfo {author} {\bibfnamefont {S.}~\bibnamefont {Uchida}}, \bibinfo {author} {\bibfnamefont {R.}~\bibnamefont {Kadono}}, \bibinfo {author}
  {\bibfnamefont {K.}~\bibnamefont {Yamada}}, \bibinfo {author} {\bibfnamefont {S.}~\bibnamefont {Tajima}}, \bibinfo {author} {\bibfnamefont {T.}~\bibnamefont {Masui}}, \bibinfo {author} {\bibfnamefont {H.}~\bibnamefont {Eisaki}}, \bibinfo {author} {\bibfnamefont {N.}~\bibnamefont {Kaneko}}, \bibinfo {author} {\bibfnamefont {M.}~\bibnamefont {Greven}},\ and\ \bibinfo {author} {\bibfnamefont {G.~D.}\ \bibnamefont {Gu}},\ }\bibfield  {title} {\bibinfo {title} {{Muon Spin Relaxation Studies of Magnetic-Field-Induced Effects in High-$T_c$ Superconductors}},\ }\href {https://doi.org/10.1103/PhysRevLett.95.157001} {\bibfield  {journal} {\bibinfo  {journal} {Physical Review Letters}\ }\textbf {\bibinfo {volume} {95}},\ \bibinfo {pages} {157001} (\bibinfo {year} {2005})}\BibitemShut {NoStop}%
\bibitem [{\citenamefont {Guguchia}\ \emph {et~al.}(2017)\citenamefont {Guguchia}, \citenamefont {Roessli}, \citenamefont {Khasanov}, \citenamefont {Amato}, \citenamefont {Pomjakushina}, \citenamefont {Conder}, \citenamefont {Uemura}, \citenamefont {Tranquada}, \citenamefont {Keller},\ and\ \citenamefont {Shengelaya}}]{17_guguchia}%
  \BibitemOpen
  \bibfield  {author} {\bibinfo {author} {\bibfnamefont {Z.}~\bibnamefont {Guguchia}}, \bibinfo {author} {\bibfnamefont {B.}~\bibnamefont {Roessli}}, \bibinfo {author} {\bibfnamefont {R.}~\bibnamefont {Khasanov}}, \bibinfo {author} {\bibfnamefont {A.}~\bibnamefont {Amato}}, \bibinfo {author} {\bibfnamefont {E.}~\bibnamefont {Pomjakushina}}, \bibinfo {author} {\bibfnamefont {K.}~\bibnamefont {Conder}}, \bibinfo {author} {\bibfnamefont {Y.~J.}\ \bibnamefont {Uemura}}, \bibinfo {author} {\bibfnamefont {J.~M.}\ \bibnamefont {Tranquada}}, \bibinfo {author} {\bibfnamefont {H.}~\bibnamefont {Keller}},\ and\ \bibinfo {author} {\bibfnamefont {A.}~\bibnamefont {Shengelaya}},\ }\bibfield  {title} {\bibinfo {title} {{Complementary response of static spin-stripe order and superconductivity to nonmagnetic impurities in cuprates}},\ }\href {https://doi.org/10.1103/PhysRevLett.119.087002} {\bibfield  {journal} {\bibinfo  {journal} {Phys. Rev. Lett.}\ }\textbf {\bibinfo {volume} {119}},\ \bibinfo {pages} {087002} (\bibinfo
  {year} {2017})}\BibitemShut {NoStop}%
\bibitem [{\citenamefont {Kojima}\ \emph {et~al.}(2003)\citenamefont {Kojima}, \citenamefont {Uchida}, \citenamefont {Fudamoto}, \citenamefont {Gat}, \citenamefont {Larkin}, \citenamefont {Uemura},\ and\ \citenamefont {Luke}}]{03_kojima}%
  \BibitemOpen
  \bibfield  {author} {\bibinfo {author} {\bibfnamefont {K.}~\bibnamefont {Kojima}}, \bibinfo {author} {\bibfnamefont {S.}~\bibnamefont {Uchida}}, \bibinfo {author} {\bibfnamefont {Y.}~\bibnamefont {Fudamoto}}, \bibinfo {author} {\bibfnamefont {I.}~\bibnamefont {Gat}}, \bibinfo {author} {\bibfnamefont {M.}~\bibnamefont {Larkin}}, \bibinfo {author} {\bibfnamefont {Y.}~\bibnamefont {Uemura}},\ and\ \bibinfo {author} {\bibfnamefont {G.}~\bibnamefont {Luke}},\ }\bibfield  {title} {\bibinfo {title} {{Superfluid density and volume fraction of static magnetism in stripe-stabilized \ch{La_{1.85−y}Eu_{y}Sr_{0.15}CuO4}}},\ }\href {https://doi.org/10.1016/S0921-4526(02)01645-9} {\bibfield  {journal} {\bibinfo  {journal} {Physica B: Condensed Matter}\ }\textbf {\bibinfo {volume} {326}},\ \bibinfo {pages} {316–320} (\bibinfo {year} {2003})}\BibitemShut {NoStop}%
\bibitem [{\citenamefont {Doiron-Leyraud}\ \emph {et~al.}(2017)\citenamefont {Doiron-Leyraud}, \citenamefont {Cyr-Choinière}, \citenamefont {Badoux}, \citenamefont {Ataei}, \citenamefont {Collignon}, \citenamefont {Gourgout}, \citenamefont {Dufour-Beauséjour}, \citenamefont {Tafti}, \citenamefont {Laliberté}, \citenamefont {Boulanger}, \citenamefont {Matusiak}, \citenamefont {Graf}, \citenamefont {Kim}, \citenamefont {Zhou}, \citenamefont {Momono}, \citenamefont {Kurosawa}, \citenamefont {Takagi},\ and\ \citenamefont {Taillefer}}]{17_doiron}%
  \BibitemOpen
  \bibfield  {author} {\bibinfo {author} {\bibfnamefont {N.}~\bibnamefont {Doiron-Leyraud}}, \bibinfo {author} {\bibfnamefont {O.}~\bibnamefont {Cyr-Choinière}}, \bibinfo {author} {\bibfnamefont {S.}~\bibnamefont {Badoux}}, \bibinfo {author} {\bibfnamefont {A.}~\bibnamefont {Ataei}}, \bibinfo {author} {\bibfnamefont {C.}~\bibnamefont {Collignon}}, \bibinfo {author} {\bibfnamefont {A.}~\bibnamefont {Gourgout}}, \bibinfo {author} {\bibfnamefont {S.}~\bibnamefont {Dufour-Beauséjour}}, \bibinfo {author} {\bibfnamefont {F.~F.}\ \bibnamefont {Tafti}}, \bibinfo {author} {\bibfnamefont {F.}~\bibnamefont {Laliberté}}, \bibinfo {author} {\bibfnamefont {M.-E.}\ \bibnamefont {Boulanger}}, \bibinfo {author} {\bibfnamefont {M.}~\bibnamefont {Matusiak}}, \bibinfo {author} {\bibfnamefont {D.}~\bibnamefont {Graf}}, \bibinfo {author} {\bibfnamefont {M.}~\bibnamefont {Kim}}, \bibinfo {author} {\bibfnamefont {J.-S.}\ \bibnamefont {Zhou}}, \bibinfo {author} {\bibfnamefont {N.}~\bibnamefont {Momono}}, \bibinfo {author}
  {\bibfnamefont {T.}~\bibnamefont {Kurosawa}}, \bibinfo {author} {\bibfnamefont {H.}~\bibnamefont {Takagi}},\ and\ \bibinfo {author} {\bibfnamefont {L.}~\bibnamefont {Taillefer}},\ }\bibfield  {title} {\bibinfo {title} {{Pseudogap phase of cuprate superconductors confined by Fermi surface topology}},\ }\href {https://doi.org/10.1038/s41467-017-02122-x} {\bibfield  {journal} {\bibinfo  {journal} {Nature Communications}\ }\textbf {\bibinfo {volume} {8}},\ \bibinfo {pages} {2044} (\bibinfo {year} {2017})}\BibitemShut {NoStop}%
\bibitem [{\citenamefont {Missiaen}\ \emph {et~al.}(2025)\citenamefont {Missiaen}, \citenamefont {Mayaffre}, \citenamefont {Krämer}, \citenamefont {Zhao}, \citenamefont {Zhou}, \citenamefont {Wu}, \citenamefont {Chen}, \citenamefont {Pyon}, \citenamefont {Takayama}, \citenamefont {Takagi}, \citenamefont {LeBoeuf},\ and\ \citenamefont {Julien}}]{25_missiaen}%
  \BibitemOpen
  \bibfield  {author} {\bibinfo {author} {\bibfnamefont {A.}~\bibnamefont {Missiaen}}, \bibinfo {author} {\bibfnamefont {H.}~\bibnamefont {Mayaffre}}, \bibinfo {author} {\bibfnamefont {S.}~\bibnamefont {Krämer}}, \bibinfo {author} {\bibfnamefont {D.}~\bibnamefont {Zhao}}, \bibinfo {author} {\bibfnamefont {Y.}~\bibnamefont {Zhou}}, \bibinfo {author} {\bibfnamefont {T.}~\bibnamefont {Wu}}, \bibinfo {author} {\bibfnamefont {X.}~\bibnamefont {Chen}}, \bibinfo {author} {\bibfnamefont {S.}~\bibnamefont {Pyon}}, \bibinfo {author} {\bibfnamefont {T.}~\bibnamefont {Takayama}}, \bibinfo {author} {\bibfnamefont {H.}~\bibnamefont {Takagi}}, \bibinfo {author} {\bibfnamefont {D.}~\bibnamefont {LeBoeuf}},\ and\ \bibinfo {author} {\bibfnamefont {M.-H.}\ \bibnamefont {Julien}},\ }\bibfield  {title} {\bibinfo {title} {{Spin-Stripe Order Tied to the Pseudogap Phase in \ch{La_{1.8-$x$}Eu_{0.2}Sr_{$x$}CuO4}}},\ }\href {https://doi.org/10.1103/PhysRevX.15.021010} {\bibfield  {journal} {\bibinfo  {journal} {Physical Review X}\
  }\textbf {\bibinfo {volume} {15}},\ \bibinfo {pages} {021010} (\bibinfo {year} {2025})}\BibitemShut {NoStop}%
\bibitem [{\citenamefont {Šimkovic}\ \emph {et~al.}(2024)\citenamefont {Šimkovic}, \citenamefont {Rossi}, \citenamefont {Georges},\ and\ \citenamefont {Ferrero}}]{24_simkovic}%
  \BibitemOpen
  \bibfield  {author} {\bibinfo {author} {\bibfnamefont {F.}~\bibnamefont {Šimkovic}}, \bibinfo {author} {\bibfnamefont {R.}~\bibnamefont {Rossi}}, \bibinfo {author} {\bibfnamefont {A.}~\bibnamefont {Georges}},\ and\ \bibinfo {author} {\bibfnamefont {M.}~\bibnamefont {Ferrero}},\ }\bibfield  {title} {\bibinfo {title} {{Origin and fate of the pseudogap in the doped Hubbard model}},\ }\href {https://doi.org/10.1126/science.ade9194} {\bibfield  {journal} {\bibinfo  {journal} {Science}\ }\textbf {\bibinfo {volume} {385}},\ \bibinfo {pages} {eade9194} (\bibinfo {year} {2024})}\BibitemShut {NoStop}%
\bibitem [{\citenamefont {Lee}(2014)}]{14_lee}%
  \BibitemOpen
  \bibfield  {author} {\bibinfo {author} {\bibfnamefont {P.~A.}\ \bibnamefont {Lee}},\ }\bibfield  {title} {\bibinfo {title} {{Amperean Pairing and the Pseudogap Phase of Cuprate Superconductors}},\ }\bibfield  {journal} {\bibinfo  {journal} {Physical Review X}\ }\textbf {\bibinfo {volume} {4}},\ \href {https://doi.org/10.1103/PhysRevX.4.031017} {10.1103/PhysRevX.4.031017} (\bibinfo {year} {2014})\BibitemShut {NoStop}%
\bibitem [{\citenamefont {Dai}\ \emph {et~al.}(2018)\citenamefont {Dai}, \citenamefont {Zhang}, \citenamefont {Senthil},\ and\ \citenamefont {Lee}}]{18_dai}%
  \BibitemOpen
  \bibfield  {author} {\bibinfo {author} {\bibfnamefont {Z.}~\bibnamefont {Dai}}, \bibinfo {author} {\bibfnamefont {Y.-H.}\ \bibnamefont {Zhang}}, \bibinfo {author} {\bibfnamefont {T.}~\bibnamefont {Senthil}},\ and\ \bibinfo {author} {\bibfnamefont {P.~A.}\ \bibnamefont {Lee}},\ }\bibfield  {title} {\bibinfo {title} {{Pair-density waves, charge-density waves, and vortices in high-$T_c$ cuprates}},\ }\href {https://doi.org/10.1103/PhysRevB.97.174511} {\bibfield  {journal} {\bibinfo  {journal} {Physical Review B}\ }\textbf {\bibinfo {volume} {97}},\ \bibinfo {pages} {174511} (\bibinfo {year} {2018})}\BibitemShut {NoStop}%
\bibitem [{\citenamefont {Lee}(2019)}]{19_lee}%
  \BibitemOpen
  \bibfield  {author} {\bibinfo {author} {\bibfnamefont {P.~A.}\ \bibnamefont {Lee}},\ }\bibfield  {title} {\bibinfo {title} {{Proposal to measure the pair field correlator of a fluctuating pair density wave}},\ }\href {https://doi.org/10.1103/PhysRevB.99.035132} {\bibfield  {journal} {\bibinfo  {journal} {Physical Review B}\ }\textbf {\bibinfo {volume} {99}},\ \bibinfo {pages} {035132} (\bibinfo {year} {2019})}\BibitemShut {NoStop}%
\bibitem [{\citenamefont {Kumagai}\ \emph {et~al.}(1994)\citenamefont {Kumagai}, \citenamefont {Kawano}, \citenamefont {Watanabe}, \citenamefont {Nishiyama},\ and\ \citenamefont {Nagamine}}]{94_kumagai}%
  \BibitemOpen
  \bibfield  {author} {\bibinfo {author} {\bibfnamefont {K.-i.}\ \bibnamefont {Kumagai}}, \bibinfo {author} {\bibfnamefont {K.}~\bibnamefont {Kawano}}, \bibinfo {author} {\bibfnamefont {I.}~\bibnamefont {Watanabe}}, \bibinfo {author} {\bibfnamefont {K.}~\bibnamefont {Nishiyama}},\ and\ \bibinfo {author} {\bibfnamefont {K.}~\bibnamefont {Nagamine}},\ }\bibfield  {title} {\bibinfo {title} {{Magnetic order and evolution of the electronic state around $x$ = 0.12 in \ch{La_{2−x}Ba_{x}CuO4} and \ch{La_{2−x}Sr_{x}CuO4}}},\ }\href {https://doi.org/10.1007/BF00730370} {\bibfield  {journal} {\bibinfo  {journal} {Journal of Superconductivity}\ }\textbf {\bibinfo {volume} {7}},\ \bibinfo {pages} {63–67} (\bibinfo {year} {1994})}\BibitemShut {NoStop}%
\bibitem [{\citenamefont {R\o{}mer}\ \emph {et~al.}(2013)\citenamefont {R\o{}mer}, \citenamefont {Chang}, \citenamefont {Christensen}, \citenamefont {Andersen}, \citenamefont {Lefmann}, \citenamefont {M\"ahler}, \citenamefont {Gavilano}, \citenamefont {Gilardi}, \citenamefont {Niedermayer}, \citenamefont {R\o{}nnow}, \citenamefont {Schneidewind}, \citenamefont {Link}, \citenamefont {Oda}, \citenamefont {Ido}, \citenamefont {Momono},\ and\ \citenamefont {Mesot}}]{13_romer}%
  \BibitemOpen
  \bibfield  {author} {\bibinfo {author} {\bibfnamefont {A.~T.}\ \bibnamefont {R\o{}mer}}, \bibinfo {author} {\bibfnamefont {J.}~\bibnamefont {Chang}}, \bibinfo {author} {\bibfnamefont {N.~B.}\ \bibnamefont {Christensen}}, \bibinfo {author} {\bibfnamefont {B.~M.}\ \bibnamefont {Andersen}}, \bibinfo {author} {\bibfnamefont {K.}~\bibnamefont {Lefmann}}, \bibinfo {author} {\bibfnamefont {L.}~\bibnamefont {M\"ahler}}, \bibinfo {author} {\bibfnamefont {J.}~\bibnamefont {Gavilano}}, \bibinfo {author} {\bibfnamefont {R.}~\bibnamefont {Gilardi}}, \bibinfo {author} {\bibfnamefont {C.}~\bibnamefont {Niedermayer}}, \bibinfo {author} {\bibfnamefont {H.~M.}\ \bibnamefont {R\o{}nnow}}, \bibinfo {author} {\bibfnamefont {A.}~\bibnamefont {Schneidewind}}, \bibinfo {author} {\bibfnamefont {P.}~\bibnamefont {Link}}, \bibinfo {author} {\bibfnamefont {M.}~\bibnamefont {Oda}}, \bibinfo {author} {\bibfnamefont {M.}~\bibnamefont {Ido}}, \bibinfo {author} {\bibfnamefont {N.}~\bibnamefont {Momono}},\ and\ \bibinfo {author}
  {\bibfnamefont {J.}~\bibnamefont {Mesot}},\ }\bibfield  {title} {\bibinfo {title} {{Glassy low-energy spin fluctuations and anisotropy gap in \ch{La_{1.88}Sr_{0.12}CuO4}}},\ }\href {https://doi.org/10.1103/PhysRevB.87.144513} {\bibfield  {journal} {\bibinfo  {journal} {Phys. Rev. B}\ }\textbf {\bibinfo {volume} {87}},\ \bibinfo {pages} {144513} (\bibinfo {year} {2013})}\BibitemShut {NoStop}%
\bibitem [{\citenamefont {Wen}\ \emph {et~al.}(2023)\citenamefont {Wen}, \citenamefont {He}, \citenamefont {Jang}, \citenamefont {Nojiri}, \citenamefont {Matsuzawa}, \citenamefont {Song}, \citenamefont {Chollet}, \citenamefont {Zhu}, \citenamefont {Liu}, \citenamefont {Fujita}, \citenamefont {Jiang}, \citenamefont {Rotundu}, \citenamefont {Kao}, \citenamefont {Jiang}, \citenamefont {Lee},\ and\ \citenamefont {Lee}}]{23_wen}%
  \BibitemOpen
  \bibfield  {author} {\bibinfo {author} {\bibfnamefont {J.-J.}\ \bibnamefont {Wen}}, \bibinfo {author} {\bibfnamefont {W.}~\bibnamefont {He}}, \bibinfo {author} {\bibfnamefont {H.}~\bibnamefont {Jang}}, \bibinfo {author} {\bibfnamefont {H.}~\bibnamefont {Nojiri}}, \bibinfo {author} {\bibfnamefont {S.}~\bibnamefont {Matsuzawa}}, \bibinfo {author} {\bibfnamefont {S.}~\bibnamefont {Song}}, \bibinfo {author} {\bibfnamefont {M.}~\bibnamefont {Chollet}}, \bibinfo {author} {\bibfnamefont {D.}~\bibnamefont {Zhu}}, \bibinfo {author} {\bibfnamefont {Y.-J.}\ \bibnamefont {Liu}}, \bibinfo {author} {\bibfnamefont {M.}~\bibnamefont {Fujita}}, \bibinfo {author} {\bibfnamefont {J.~M.}\ \bibnamefont {Jiang}}, \bibinfo {author} {\bibfnamefont {C.~R.}\ \bibnamefont {Rotundu}}, \bibinfo {author} {\bibfnamefont {C.-C.}\ \bibnamefont {Kao}}, \bibinfo {author} {\bibfnamefont {H.-C.}\ \bibnamefont {Jiang}}, \bibinfo {author} {\bibfnamefont {J.-S.}\ \bibnamefont {Lee}},\ and\ \bibinfo {author} {\bibfnamefont {Y.~S.}\ \bibnamefont
  {Lee}},\ }\bibfield  {title} {\bibinfo {title} {{Enhanced charge density wave with mobile superconducting vortices in \ch{La_{1.885}Sr_{0.115}CuO4}}},\ }\href {https://doi.org/10.1038/s41467-023-36203-x} {\bibfield  {journal} {\bibinfo  {journal} {Nat. Commun.}\ }\textbf {\bibinfo {volume} {14}},\ \bibinfo {pages} {733} (\bibinfo {year} {2023})}\BibitemShut {NoStop}%
\bibitem [{\citenamefont {Chen}\ \emph {et~al.}(2024)\citenamefont {Chen}, \citenamefont {Huang}, \citenamefont {Ma}, \citenamefont {Smith}, \citenamefont {Sharron}, \citenamefont {Aczel}, \citenamefont {Tian},\ and\ \citenamefont {Gaulin}}]{24_chen}%
  \BibitemOpen
  \bibfield  {author} {\bibinfo {author} {\bibfnamefont {Q.}~\bibnamefont {Chen}}, \bibinfo {author} {\bibfnamefont {S.~H.-Y.}\ \bibnamefont {Huang}}, \bibinfo {author} {\bibfnamefont {Q.}~\bibnamefont {Ma}}, \bibinfo {author} {\bibfnamefont {E.~M.}\ \bibnamefont {Smith}}, \bibinfo {author} {\bibfnamefont {H.}~\bibnamefont {Sharron}}, \bibinfo {author} {\bibfnamefont {A.~A.}\ \bibnamefont {Aczel}}, \bibinfo {author} {\bibfnamefont {W.}~\bibnamefont {Tian}},\ and\ \bibinfo {author} {\bibfnamefont {B.~D.}\ \bibnamefont {Gaulin}},\ }\bibfield  {title} {\bibinfo {title} {{Random fields from quenched disorder in an archetype for correlated electrons: The parallel spin stripe phase of \ch{La_{1.6-$x$}Nd_{0.4}Sr_{$x$}CuO4} at the 1/8 anomaly}},\ }\href {https://doi.org/10.1103/PhysRevB.109.134411} {\bibfield  {journal} {\bibinfo  {journal} {Phys. Rev. B}\ }\textbf {\bibinfo {volume} {109}},\ \bibinfo {pages} {134411} (\bibinfo {year} {2024})}\BibitemShut {NoStop}%
\bibitem [{\citenamefont {Tranquada}\ \emph {et~al.}(2008)\citenamefont {Tranquada}, \citenamefont {Gu}, \citenamefont {Hücker}, \citenamefont {Jie}, \citenamefont {Kang}, \citenamefont {Klingeler}, \citenamefont {Li}, \citenamefont {Tristan}, \citenamefont {Wen}, \citenamefont {Xu}, \citenamefont {Xu}, \citenamefont {Zhou},\ and\ \citenamefont {v.~Zimmermann}}]{08_tranquada}%
  \BibitemOpen
  \bibfield  {author} {\bibinfo {author} {\bibfnamefont {J.~M.}\ \bibnamefont {Tranquada}}, \bibinfo {author} {\bibfnamefont {G.~D.}\ \bibnamefont {Gu}}, \bibinfo {author} {\bibfnamefont {M.}~\bibnamefont {Hücker}}, \bibinfo {author} {\bibfnamefont {Q.}~\bibnamefont {Jie}}, \bibinfo {author} {\bibfnamefont {H.-J.}\ \bibnamefont {Kang}}, \bibinfo {author} {\bibfnamefont {R.}~\bibnamefont {Klingeler}}, \bibinfo {author} {\bibfnamefont {Q.}~\bibnamefont {Li}}, \bibinfo {author} {\bibfnamefont {N.}~\bibnamefont {Tristan}}, \bibinfo {author} {\bibfnamefont {J.~S.}\ \bibnamefont {Wen}}, \bibinfo {author} {\bibfnamefont {G.~Y.}\ \bibnamefont {Xu}}, \bibinfo {author} {\bibfnamefont {Z.~J.}\ \bibnamefont {Xu}}, \bibinfo {author} {\bibfnamefont {J.}~\bibnamefont {Zhou}},\ and\ \bibinfo {author} {\bibfnamefont {M.}~\bibnamefont {v.~Zimmermann}},\ }\bibfield  {title} {\bibinfo {title} {{Evidence for unusual superconducting correlations coexisting with stripe order in \ch{La_{1.875}Ba_{0.125}CuO4}}},\ }\href
  {https://doi.org/10.1103/PhysRevB.78.174529} {\bibfield  {journal} {\bibinfo  {journal} {Physical Review B}\ }\textbf {\bibinfo {volume} {78}},\ \bibinfo {pages} {174529} (\bibinfo {year} {2008})}\BibitemShut {NoStop}%
\bibitem [{\citenamefont {Wen}\ \emph {et~al.}(2019)\citenamefont {Wen}, \citenamefont {Huang}, \citenamefont {Lee}, \citenamefont {Jang}, \citenamefont {Knight}, \citenamefont {Lee}, \citenamefont {Fujita}, \citenamefont {Suzuki}, \citenamefont {Asano}, \citenamefont {Kivelson}, \citenamefont {Kao},\ and\ \citenamefont {Lee}}]{19_wen}%
  \BibitemOpen
  \bibfield  {author} {\bibinfo {author} {\bibfnamefont {J.-J.}\ \bibnamefont {Wen}}, \bibinfo {author} {\bibfnamefont {H.}~\bibnamefont {Huang}}, \bibinfo {author} {\bibfnamefont {S.-J.}\ \bibnamefont {Lee}}, \bibinfo {author} {\bibfnamefont {H.}~\bibnamefont {Jang}}, \bibinfo {author} {\bibfnamefont {J.}~\bibnamefont {Knight}}, \bibinfo {author} {\bibfnamefont {Y.~S.}\ \bibnamefont {Lee}}, \bibinfo {author} {\bibfnamefont {M.}~\bibnamefont {Fujita}}, \bibinfo {author} {\bibfnamefont {K.~M.}\ \bibnamefont {Suzuki}}, \bibinfo {author} {\bibfnamefont {S.}~\bibnamefont {Asano}}, \bibinfo {author} {\bibfnamefont {S.~A.}\ \bibnamefont {Kivelson}}, \bibinfo {author} {\bibfnamefont {C.-C.}\ \bibnamefont {Kao}},\ and\ \bibinfo {author} {\bibfnamefont {J.-S.}\ \bibnamefont {Lee}},\ }\bibfield  {title} {\bibinfo {title} {{Observation of two types of charge-density-wave orders in superconducting \ch{La_{2-$x$}Sr_{$x$}CuO4}}},\ }\href {https://doi.org/10.1038/s41467-019-11167-z} {\bibfield  {journal} {\bibinfo
  {journal} {Nat. Commun.}\ }\textbf {\bibinfo {volume} {10}},\ \bibinfo {pages} {3269} (\bibinfo {year} {2019})}\BibitemShut {NoStop}%
\bibitem [{\citenamefont {Luke}\ \emph {et~al.}(1991)\citenamefont {Luke}, \citenamefont {Le}, \citenamefont {Sternlieb}, \citenamefont {Wu}, \citenamefont {Uemura}, \citenamefont {Brewer}, \citenamefont {Riseman}, \citenamefont {Ishibashi},\ and\ \citenamefont {Uchida}}]{91_luke}%
  \BibitemOpen
  \bibfield  {author} {\bibinfo {author} {\bibfnamefont {G.~M.}\ \bibnamefont {Luke}}, \bibinfo {author} {\bibfnamefont {L.~P.}\ \bibnamefont {Le}}, \bibinfo {author} {\bibfnamefont {B.~J.}\ \bibnamefont {Sternlieb}}, \bibinfo {author} {\bibfnamefont {W.~D.}\ \bibnamefont {Wu}}, \bibinfo {author} {\bibfnamefont {Y.~J.}\ \bibnamefont {Uemura}}, \bibinfo {author} {\bibfnamefont {J.~H.}\ \bibnamefont {Brewer}}, \bibinfo {author} {\bibfnamefont {T.~M.}\ \bibnamefont {Riseman}}, \bibinfo {author} {\bibfnamefont {S.}~\bibnamefont {Ishibashi}},\ and\ \bibinfo {author} {\bibfnamefont {S.}~\bibnamefont {Uchida}},\ }\bibfield  {title} {\bibinfo {title} {{Static magnetic order in \ch{La_{1.875}Ba_{0.125}CuO4}}},\ }\href@noop {} {\bibfield  {journal} {\bibinfo  {journal} {Physica C: Superconductivity}\ }\textbf {\bibinfo {volume} {185}},\ \bibinfo {pages} {1175} (\bibinfo {year} {1991})}\BibitemShut {NoStop}%
\bibitem [{\citenamefont {Lozano}\ \emph {et~al.}(2022)\citenamefont {Lozano}, \citenamefont {Ren}, \citenamefont {Gu}, \citenamefont {Tsvelik}, \citenamefont {Tranquada},\ and\ \citenamefont {Li}}]{22_lozano}%
  \BibitemOpen
  \bibfield  {author} {\bibinfo {author} {\bibfnamefont {P.~M.}\ \bibnamefont {Lozano}}, \bibinfo {author} {\bibfnamefont {T.}~\bibnamefont {Ren}}, \bibinfo {author} {\bibfnamefont {G.~D.}\ \bibnamefont {Gu}}, \bibinfo {author} {\bibfnamefont {A.~M.}\ \bibnamefont {Tsvelik}}, \bibinfo {author} {\bibfnamefont {J.~M.}\ \bibnamefont {Tranquada}},\ and\ \bibinfo {author} {\bibfnamefont {Q.}~\bibnamefont {Li}},\ }\bibfield  {title} {\bibinfo {title} {{Testing for pair density wave order in \ch{La_{1.875}Ba_{0.125}CuO4}}},\ }\href {https://doi.org/10.1103/PhysRevB.106.174510} {\bibfield  {journal} {\bibinfo  {journal} {Physical Review B}\ }\textbf {\bibinfo {volume} {106}},\ \bibinfo {pages} {174510} (\bibinfo {year} {2022})}\BibitemShut {NoStop}%
\bibitem [{\citenamefont {Ren}\ \emph {et~al.}(2023)\citenamefont {Ren}, \citenamefont {Lozano}, \citenamefont {Li}, \citenamefont {Gu},\ and\ \citenamefont {Tsvelik}}]{23_ren}%
  \BibitemOpen
  \bibfield  {author} {\bibinfo {author} {\bibfnamefont {T.}~\bibnamefont {Ren}}, \bibinfo {author} {\bibfnamefont {P.~M.}\ \bibnamefont {Lozano}}, \bibinfo {author} {\bibfnamefont {Q.}~\bibnamefont {Li}}, \bibinfo {author} {\bibfnamefont {G.}~\bibnamefont {Gu}},\ and\ \bibinfo {author} {\bibfnamefont {A.~M.}\ \bibnamefont {Tsvelik}},\ }\bibfield  {title} {\bibinfo {title} {{Two types of superconducting pairs in stripe-ordered LBCO ($x$ = 1/8): Evidence from resistivity measurements}},\ }\href {https://doi.org/10.1103/PhysRevB.107.085118} {\bibfield  {journal} {\bibinfo  {journal} {Physical Review B}\ }\textbf {\bibinfo {volume} {107}},\ \bibinfo {pages} {085118} (\bibinfo {year} {2023})}\BibitemShut {NoStop}%
\bibitem [{\citenamefont {Ding}\ \emph {et~al.}(2008)\citenamefont {Ding}, \citenamefont {Xiang}, \citenamefont {Zhang}, \citenamefont {Liu},\ and\ \citenamefont {Li}}]{08_ding}%
  \BibitemOpen
  \bibfield  {author} {\bibinfo {author} {\bibfnamefont {J.~F.}\ \bibnamefont {Ding}}, \bibinfo {author} {\bibfnamefont {X.~Q.}\ \bibnamefont {Xiang}}, \bibinfo {author} {\bibfnamefont {Y.~Q.}\ \bibnamefont {Zhang}}, \bibinfo {author} {\bibfnamefont {H.}~\bibnamefont {Liu}},\ and\ \bibinfo {author} {\bibfnamefont {X.~G.}\ \bibnamefont {Li}},\ }\bibfield  {title} {\bibinfo {title} {{Two-dimensional superconductivity in stripe-ordered \ch{La_{1.6 − x}Nd_{0.4}Sr_{x}CuO4} single crystals}},\ }\href {https://doi.org/10.1103/PhysRevB.77.214524} {\bibfield  {journal} {\bibinfo  {journal} {Physical Review B}\ }\textbf {\bibinfo {volume} {77}},\ \bibinfo {pages} {214524} (\bibinfo {year} {2008})}\BibitemShut {NoStop}%
\bibitem [{\citenamefont {Shi}\ \emph {et~al.}(2020{\natexlab{a}})\citenamefont {Shi}, \citenamefont {Baity}, \citenamefont {Terzic}, \citenamefont {Sasagawa},\ and\ \citenamefont {Popović}}]{20_shi_NC}%
  \BibitemOpen
  \bibfield  {author} {\bibinfo {author} {\bibfnamefont {Z.}~\bibnamefont {Shi}}, \bibinfo {author} {\bibfnamefont {P.~G.}\ \bibnamefont {Baity}}, \bibinfo {author} {\bibfnamefont {J.}~\bibnamefont {Terzic}}, \bibinfo {author} {\bibfnamefont {T.}~\bibnamefont {Sasagawa}},\ and\ \bibinfo {author} {\bibfnamefont {D.}~\bibnamefont {Popović}},\ }\bibfield  {title} {\bibinfo {title} {Pair density wave at high magnetic fields in cuprates with charge and spin orders},\ }\href {https://doi.org/10.1038/s41467-020-17138-z} {\bibfield  {journal} {\bibinfo  {journal} {Nature Communications}\ }\textbf {\bibinfo {volume} {11}},\ \bibinfo {pages} {3323} (\bibinfo {year} {2020}{\natexlab{a}})}\BibitemShut {NoStop}%
\bibitem [{\citenamefont {Shi}\ \emph {et~al.}(2020{\natexlab{b}})\citenamefont {Shi}, \citenamefont {Baity}, \citenamefont {Sasagawa},\ and\ \citenamefont {Popović}}]{20_shi_SA}%
  \BibitemOpen
  \bibfield  {author} {\bibinfo {author} {\bibfnamefont {Z.}~\bibnamefont {Shi}}, \bibinfo {author} {\bibfnamefont {P.~G.}\ \bibnamefont {Baity}}, \bibinfo {author} {\bibfnamefont {T.}~\bibnamefont {Sasagawa}},\ and\ \bibinfo {author} {\bibfnamefont {D.}~\bibnamefont {Popović}},\ }\bibfield  {title} {\bibinfo {title} {Vortex phase diagram and the normal state of cuprates with charge and spin orders},\ }\href {https://doi.org/10.1126/sciadv.aay8946} {\bibfield  {journal} {\bibinfo  {journal} {Science Advances}\ }\textbf {\bibinfo {volume} {6}},\ \bibinfo {pages} {eaay8946} (\bibinfo {year} {2020}{\natexlab{b}})}\BibitemShut {NoStop}%
\end{thebibliography}%

\end{document}